\documentclass[fleqn,usenatbib]{mnras}

\usepackage{newtxtext,newtxmath}

\usepackage[T1]{fontenc}

\DeclareRobustCommand{\VAN}[3]{#2}
\let\VANthebibliography\thebibliography
\def\thebibliography{\DeclareRobustCommand{\VAN}[3]{##3}\VANthebibliography}
\defcitealias{Eltvedt2023}{Paper~I}

\usepackage{graphicx}	
\usepackage{amsmath}	

\title[VST ATLAS QSO SURVEY II]{The VST ATLAS Quasar Survey II: Halo mass profiles of galaxies, LRGs and galaxy clusters via quasar and CMB lensing}

\author[A. M. Eltvedt et al.]
{Alice M. Eltvedt,$^{1}$\thanks{E-mail: aeltvedt@alumni.princeton.edu}
T. Shanks,$^{1}$\thanks{E-mail: tom.shanks@durham.ac.uk}
N. Metcalfe,$^{1}$
B. Ansarinejad,$^{2}$
L.F. Barrientos,$^{3}$\newauthor
D.N.A. Murphy$^{3,4}$
and D.M. Alexander$^{1}$
\\
$^{1}$Centre for Extragalactic Astronomy, Department of Physics, Durham University, South Road, Durham, DH1 3LE, UK\\
$^{2}$ School of Physics, University of Melbourne, Parkville, VIC 3010, Australia\\
$^{3}$Instituto de Astrofisica, Facultad de Fisica, Pontificia Universidad Catolica de Chile, Santiago, Chile\\
$^{4}$Institute of Astronomy, University of Cambridge, Madingley Road, Cambridge CB3 0HA, UK \\
}

\date{Accepted XXX. Received YYY; in original form ZZZ}

\pubyear{2024}

\begin{document}
\label{firstpage}
\pagerange{\pageref{firstpage}--\pageref{lastpage}}
\maketitle

\begin{abstract}

We cross-correlate a low-contamination subset of the VST ATLAS $g<22.5$
quasar catalogue with $g<21.5$ galaxy clusters, $r<21$ galaxies and
$r<19.5$ Luminous Red Galaxies (LRGs) to  probe their halo mass profiles via quasar
magnification bias caused by weak lensing. In the case of galaxy
clusters we find that at small scales their mass profiles are well
fitted by Navarro, Frenk and White (NFW) models with masses within the expected range. For the galaxies, we find consistency with previous SDSS-based results  for the galaxy-quasar cross-correlation and the galaxy auto-correlation functions.
Disagreement as to whether the cross-correlation results are in tension
with $\Lambda$CDM appears due to different assumptions as to whether
galaxies  trace mass. We conclude that halo occupation distribution (HOD) models fit the galaxy - quasar
lensing results better than models where galaxies trace the mass. We
further test the cluster and galaxy HOD models in the 2-halo range using the
{\it Planck} Cosmic Microwave Background (CMB) lensing map, finding that the cross-correlation with both
the poorest clusters and the galaxies may be marginally  over-predicted
by the above HOD models. Finally, we measure the magnification bias of
LRGs using both quasar and CMB lensing and find that the  observed
quasar lensing amplitude may be $\approx2\times$ too high and, on larger
scales, the CMB lensing amplitude may be too low to be explained by a
standard LRG HOD model.

\end{abstract}

\begin{keywords}
 quasars: general - galaxies: general - galaxies: clusters: general - dark matter
\end{keywords}



\section{Introduction}

The detection of the accelerated expansion of the universe (e.g. \citealt{Riess1998, Perlmutter1999}) as well as the existence of dark matter (\citealt{Zwicky1933, Rubin1977}), as necessitated by the currently accepted $\Lambda$ Cold Dark Matter ($\Lambda$CDM) model of the universe, has made the undisputed determination and understanding of these phenomena a main goal of modern astrophysics. The existence of dark energy is one possibility to explain the accelerated expansion of our universe within the framework of Einstein's theory of general relativity (e.g. \citealt{RP1988}) and dark matter is needed to explain observations of the clustering of structures (e.g. \citealt{Peebles1980}). In terms of understanding the nature of dark matter, gravitational lensing analyses  are clearly of prime interest (e.g. \citealt{N1985,KS1993, Kaiser1998, Myers2003}). 

The lensing magnification of background objects by large scale structures can also provide constraints on the cosmological parameters, especially the matter density of the universe ($\Omega_M$) and the "clumpiness" ($\sigma_8$). Galaxy-quasar cross-correlation studies have been conducted since  \cite{SP1979} detected a possible quasar excess around Lick catalogue galaxies (see also  \citealt{BFS1988}). More recently, works by \cite{Myers2003}, \cite{Myers2005} and \cite{MS2007} have used background 2QZ \citep{Croom2005} quasars to detect the effect of galaxy and galaxy cluster lensing and \cite{Scranton2005} have performed such lensing analyses using photo-z selected quasars from the Sloan Digital Sky Survey (SDSS). \cite{Myers2003} and \cite{MS2007} found a higher than expected amplitude of lensing magnification bias based on simple $\Omega_m=0.3$ models that assumed galaxies traced the mass, and suggested there may be  inconsistency with the standard $\Lambda$CDM model. However, \cite{Scranton2005} argued conversely that their SDSS results were compatible with the standard $\Lambda$CDM model. \cite{Menard2010} confirmed these findings on the full SDSS imaging catalogue (while also detecting a sub-dominant contribution from galactic dust absorption to the cross-correlation functions). Here, we perform a weak gravitational lensing analysis through a cross-correlation of background quasars and foreground galaxies and galaxy clusters using the VST ATLAS Quasar Catalogue (see \citealt{Eltvedt2023} hereafter \citetalias{Eltvedt2023}) to provide independent new data to further address the reasons for this apparent discrepancy.  

The lensing mentioned above is defined as the gravitational deflection of photons around large masses, which causes a magnification of background sources (e.g. \citealt{Narayan1989}). This "magnification bias"  causes the background objects to appear brighter than they actually are while reducing the apparent solid angle behind the foreground objects, causing an increase in QSO density at bright QSO magnitudes where the slope of their number count is steeper and a decrease at fainter magnitudes where their number counts are flatter. Here we present our results, their interpretation and any implications for the cosmological model. We show that an anti-correlation is detected at faint quasar magnitudes and a positive correlation at detected at bright magnitudes as predicted by lensing. Through this cross-correlation we will be able to test (HOD) models and their assumed mass profiles over a wide range of halo masses.  We shall further apply these quasar lensing  analyses to galaxy cluster and LRG samples. 

As an independent alternative to quasar lensing we shall also exploit  CMB lensing (e.g. \citealt{BS1987,Seljak1996}) by cross-correlating our ATLAS galaxy cluster, galaxy and LRG catalogues with the lensing maps of the CMB supplied by \cite{Planck2018} to  measure their  halo profiles and fit HODs as above. Here we shall follow e.g. \cite{BS1987,Seljak1996}  and then more recently e.g. \cite{Krolewski2020, Krolewski2021}. We note that the resolution of the {\it Planck} lensing map is $\approx6'$, giving  information extending to larger angular scales than quasar lensing, while still allowing us to make direct comparisons between these two at intermediate scales in the 1-2 halo  regime at $\approx1h^{-1}$ Mpc. \cite{Krolewski2020, Krolewski2021} measure LRG-CMB lensing by using  unWISE W1 and W2 bands \citep{unWISE2019} to select samples of LRGs at $z=0.6, 1,1, 1.4$ to produce cross-power spectra with the {\it Planck} lensing maps. Their main interest is to measure cosmological parameters and so they confine their studies to large scales, $60'-900'$, whereas we complement the CMB-lensing with the small-scale QSO lensing to estimate halo mass profiles out to scales of $\approx0.'3-60'$, corresponding to $\approx0.1-10$h$^{-1}$ Mpc at our LRG average redshift of $z\approx0.26$. In Eltvedt et al (2024) (hereafter Paper 3) we shall also use CMB lensing to measure the halo mass profiles of our $z\approx1.7$ QSOs themselves, following in particular the work of \cite{Geach2019, Han2019, P22, P23}.

The outline of this paper is as follows. Section ~\ref{sec:QSOgalclust_lensing} and \ref{sec:QSOgal_xcorr} describes the cross-correlation of ATLAS selected foreground galaxy clusters and galaxies respectively with our quasar catalogue and the {\it Planck} CMB lensing map.  We introduce HOD models in Section~\ref{sec:HOD_models} and fit these to the quasar+CMB lensing results for the galaxy clusters and the galaxies, and also the galaxy autocorrelation function. To address the possibility that the 1-halo term is less well fitted to the galaxy cross-correlations, in Section ~\ref{sec:lrg_hod_model} we also perform quasar and CMB lensing cross-correlations of foreground LRGs and fit HOD models that are also tested against the LRG auto-correlation function. We discuss our results in Section~\ref{sec:conclusions}. Throughout, we assume a standard, spatially flat, cosmology with $\Omega_m=0.3$ and  a Hubble constant assumed to be 100 h km s$^{-1}$ Mpc$^{-1}$, with h=0.7 unless otherwise stated.

\section{Data Catalogues}
\label{sec:xcorr_catalogues}

\subsection{Quasar Sample}
\label{sec:xcorr_QSO_sample}

The VST-ATLAS quasar catalogue described in \citetalias{Eltvedt2023} has a certain amount of stellar and galaxy contamination, an inevitable consequence of requiring high quasar completeness. To perform these weak lensing analyses we use a more conservative, point-source only selection of our quasar catalogue to reduce galaxy contamination as well as possible overlap in the galaxy and quasar catalogues. We use the quasar candidate catalogue with the $ugri+giW1W2$ cuts described in Section 4 of \citetalias{Eltvedt2023}. We then further restrict this  point-source candidate selection to $17<g<22$. 

\begin{figure}
	\centering
 	\includegraphics[width=\columnwidth]{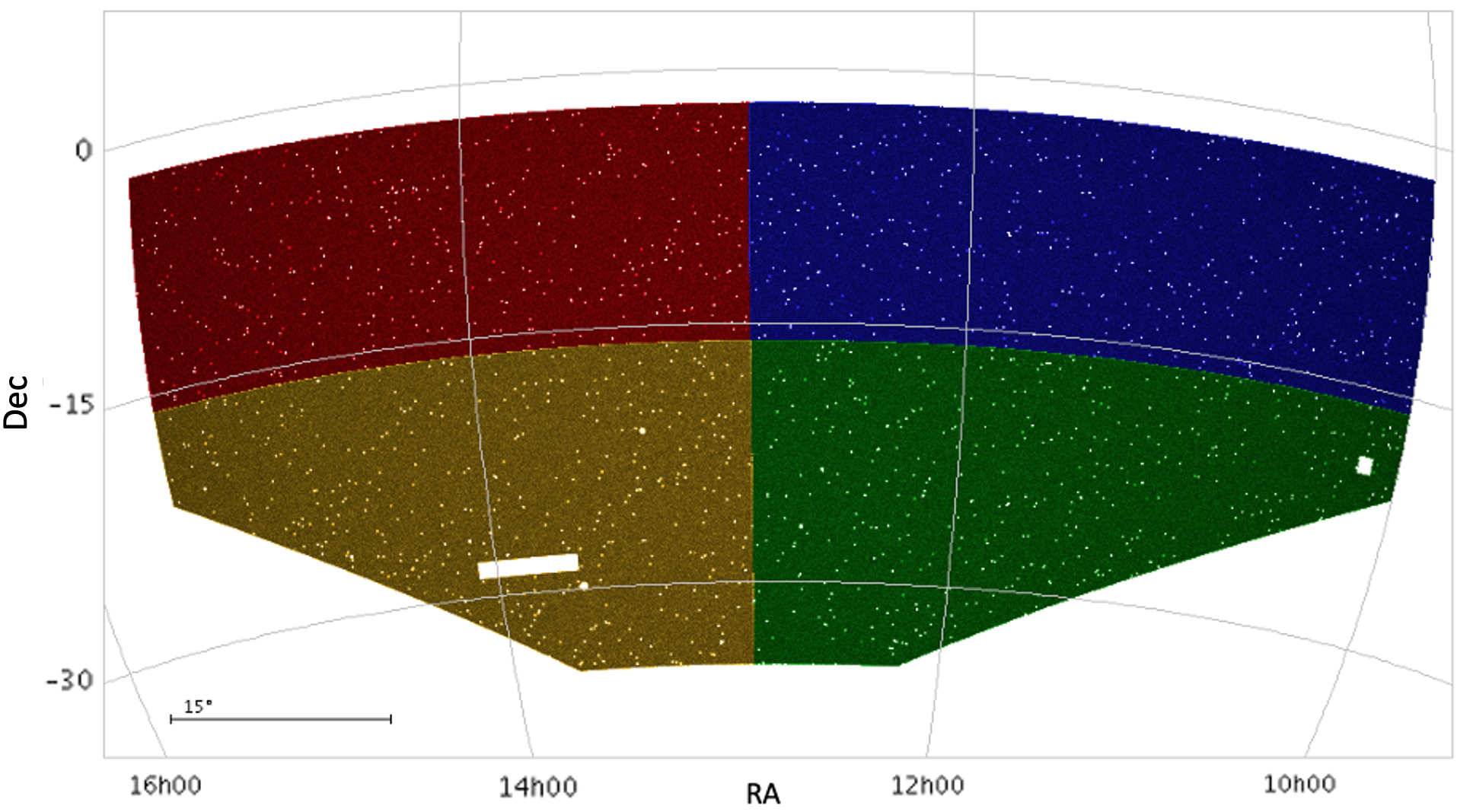}
    \includegraphics[width=\columnwidth]{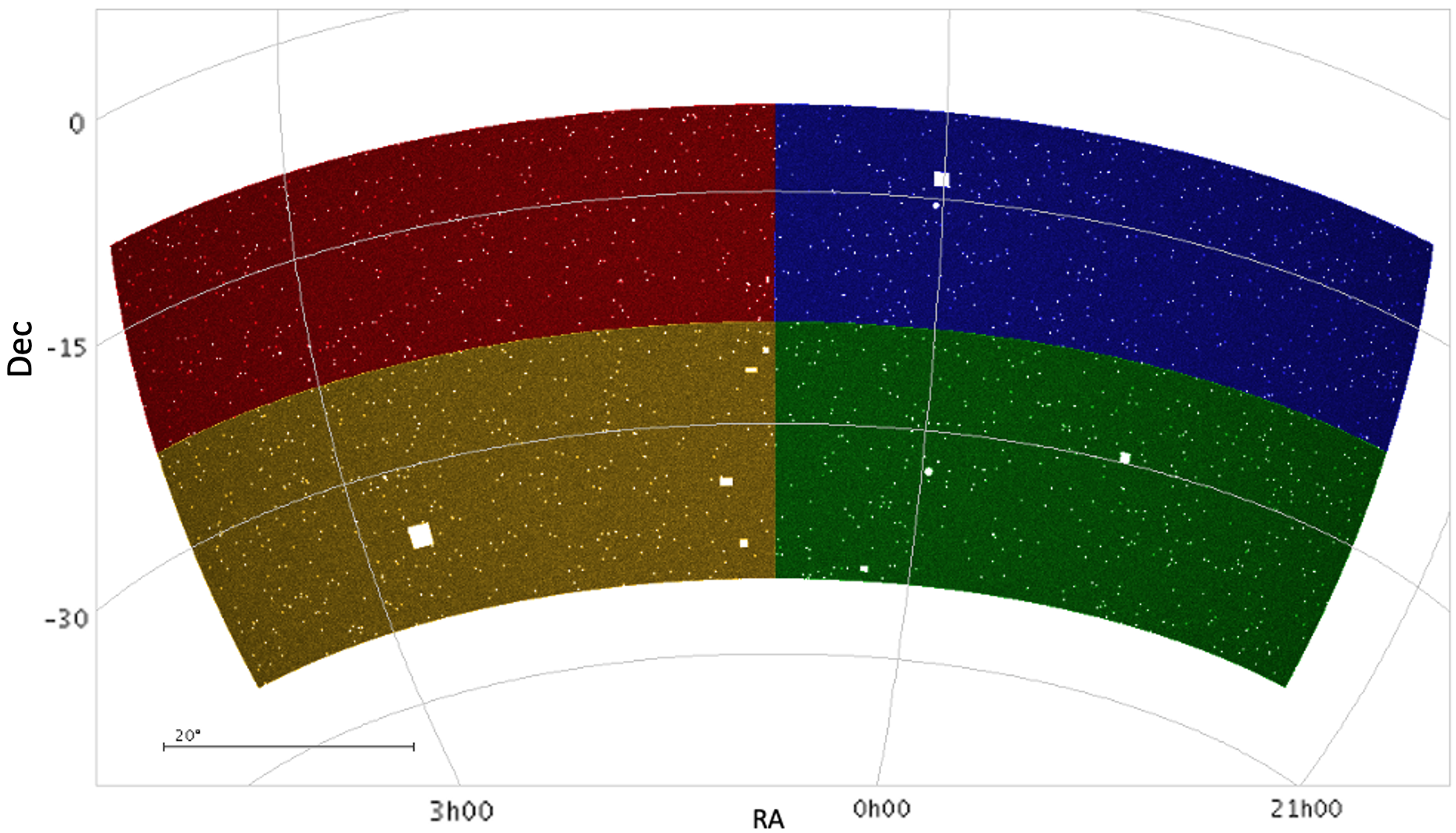}
	\caption[Maps of our random catalogues in the NGC (upper panel) and SGC (lower panel), split into 8 equal area regions to calculate errors from field-to-field variations.]{Maps of our random catalogue in the NGC (upper panel) and SGC (lower panel), covering the same areas as our quasar candidate catalogue. The catalogues are  split into 8 approximately equal area regions to calculate errors from field-to-field variations. We have masked out Tycho stars, globular clusters, nearby dwarf galaxies and areas that are underdense due to poor observing conditions. These are left as white areas in the maps. }
	\label{fig:NGC_racut}
\end{figure}

Following an analysis of preliminary spectroscopically confirmed QSOs, we also restrict this sample to $-0.25<(g-r)<0.4$, $(u-g)<0.55$, $(r-W1)<5$, and require $(W1-W2)>0.4$, again to reduce the possibility of galaxy contamination in our sample. Of this more conservative selection, we only consider quasar candidates with photometric redshifts $z>1$ to prevent overlap in real space of quasar and galaxy samples, using results from the ANNz2 photometric redshift estimation. We also mask areas around Tycho stars to $V_T<12.5$ following the method of \cite{Ansarinejad2023}. Also masked are globular clusters and dwarf galaxies as well as a few areas with poor photometry. These selections result in a total of $204264$ objects giving us a quasar candidate sky density of $44$deg$^{-2}$. The QSO distribution can be seen in Fig. 2 of Paper 3 and the QSO (and galaxy) masked random catalogue  is  shown in Fig.~\ref{fig:NGC_racut} (see Section \ref{sec:CUTE}).


\subsection{Galaxy Cluster Sample}
\label{sec:xcorr_galclust_sample}

We use the VST ATLAS Southern Galaxy Cluster Catalogue \citep{Ansarinejad2023} to perform the angular cross-correlations between foreground galaxy clusters and background quasars. The galaxy groups and clusters in this catalogue were selected using VST ATLAS optical photometry in the $griz$ bands using the ORCA cluster detection algorithm. The ORCA cluster detection algorithm \citep{ORCA2012} finds similarities in galaxy colours and regions with a high projected surface density and then uses the friends-of-friends technique to determine galaxy clusters and  groups.  The selection criteria are  described in full by \cite{Ansarinejad2023}. This cluster catalogue overlaps the full $\sim 4700$ deg$^2$ area of our VST-ATLAS quasar survey to a depth of $r_{Kron}<21$. \\

We introduce the same Tycho stars and globular cluster mask as in our QSO catalogue. This galaxy cluster catalogue is then divided into clusters with $5$ or more members ($n>5$) and clusters with $40$ or more members ($n>40$). The resulting $n\geq5$ catalogue has $N_g=386268$ galaxies, with a galaxy cluster member sky density of $82.18$deg$^{-2}$ and a cluster sky density of 6.54deg$^{-2}$. The $n\geq40$ catalogue has $N_g=60210$ galaxies, with a galaxy cluster member sky density of $12.81$deg$^{-2}$ and a cluster sky density of 0.19deg$^{-2}$.



\begin{figure}
	\centering
	\includegraphics[width=\columnwidth]{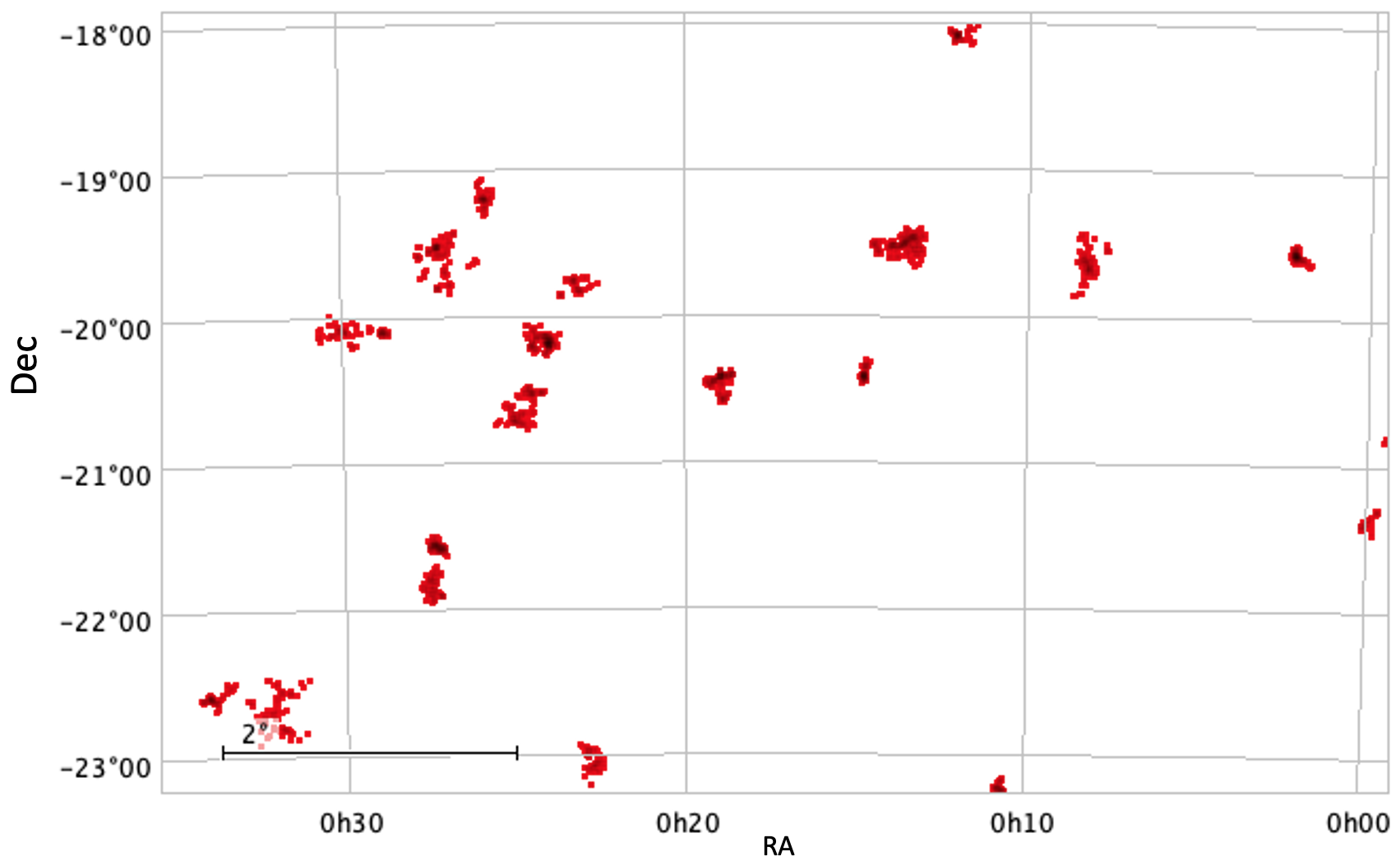}
	\caption[Sky map of defined $n\geq14$ galaxy clusters in a section of the SGC]{Sky map of defined $n\geq40$ galaxy clusters in a section of the SGC with each point corresponding to a galaxy.  }
	\label{fig:galcluster_skymap}
\end{figure}

Fig.~\ref{fig:galcluster_skymap} shows a patch of sky in the SGC from our $n\geq40$ galaxy group sample. The cross-correlations between the galaxy cluster and quasar catalogues are performed between quasars and individual members of each galaxy cluster rather than the center of the clusters. Therefore, the larger clusters are weighted more heavily.\\

\subsection{Galaxy Sample}
\label{sec:xcorr_gal_sample}

To perform the cross-correlation analyses of our quasar candidate catalogue and individual galaxies, we also generate galaxy catalogues from the VST ATLAS data using the same star/galaxy separation as for our QSO sample. To provide an accurate comparison to the work done on SDSS data by \cite{Scranton2005}, we require the galaxies to have detections to $r_{sdss}<21$, using a $0.15$ mag offset to convert from ATLAS Kron $r$ to the total $r$-band SDSS magnitudes, ie $r_{sdss}=r_{Kron}-0.15$, as suggested by previous authors (e.g. \citealt{Kron1980, Metcalfe1991, Shanks2015}).\footnote{From this point, we shall refer  to $r_{sdss}$ as $r_{Kron}$.} We use the same Tycho stars and globular cluster mask for all of our catalogues. 


\subsection{Luminous Red Galaxy Sample}
\label{sec:xcorr_lrg_sample}

We perform cross correlation with LRGs to test halo occupation models in Section \ref{sec:lrg_hod_model}. To do this, we create a catalogue of LRGs based on the "Cut 1" $z<0.4$ selection shown in Figure 3 of \cite{Eisenstein2001}, who get an LRG sky density of $14.3$ deg$^{-2}$. Applying their selections on our galaxy catalogue as described in \ref{sec:xcorr_gal_sample}, we get a sample with a sky density of $9.3$ deg$^{-2}$. As this is lower than the $14.3$deg$^{-2}$ sky density, we adjust the selection slightly from $r_{KRON}<19.2$ and $r_{KRON}<12.38+2.33*(g-r)+4*(r-i)$ to $r_{KRON}<19.5$ and $r_{KRON}<12.68+2.33*(g-r)+4*(r-i)$ to increase the density of LRGs we are getting to $16$ deg$^{-2}$, which is as close to 14.3 deg$^{-2}$ as can be achieved to 0.1mag accuracy in the magnitude limit.

\subsection{Star Control Sample}
\label{sec:xcorr_star_sample}

We create a subset of stars to check the signal of our cross-correlations between galaxy clusters and quasars, galaxies and quasars, and finally LRGs and quasars. We select stars away from the $W1$ limit by selecting stars in the same $r-W1$ range as the QSOs (i.e. $3<r-W1<5$ and $1<g-r<1.4$, see Fig. 7 of \cite{Eltvedt2023}), as we noticed that stars were being lost due to potential systematic effects, such as sky subtraction, near the W1 limit. We also go to the brighter limit of $g<21$, than the $g<22$ limit of our VST ATLAS catalogue to decrease potential contamination, creating a control sample that is as well positioned as possible to check our work. \\

Previously, when we selected star control samples in the ranges $1<r-W1<3$ and $0.3<g-r<1$ i.e. including fainter objects  in W1 than existed in the QSO sample, we found anomalies where cross-correlation of galaxies and stars showed unexpected anti-correlation. This anti-correlation appeared to increase with galaxy apparent brightness. We also found that this anti-correlation was more evident in star samples that relied on stars selected  at the faintest  W1 and W2 NEO7 magnitudes. The effect was reduced, but still not eliminated, when DECALS DR10 "forced" W1 and W2 photometry was used instead of NEO7. We hypothesize that there may be a sky subtraction bias in W1 in the vicinity of a bright galaxy where the sky brightness may be over-estimated. The effect was particularly evident in stars selected in $grW1$ to lie at $r-W1<2$. This selection is otherwise optimal in avoiding galaxy contamination (see \citetalias{Eltvedt2023}) but since our QSO samples reach $g\approx r\sim 22$ this means the equivalent star sample reaches $W1\sim21$ compared to a NEO7 limit of $W1\sim20$  so these samples suffer high incompleteness and will be more prone to the sky subtraction issue postulated above. When a control star sample with an $r-W1$ distribution more similar to the QSOs was used  (i.e. $1<g-r<1.4$ and $3<r-W1<5$), this anti-correlation reduced significantly. We considered the possibility that galaxy contamination in this star sample  might also contribute to this reduction. However, simple $g<22.5$ star samples with no colour selection also gave no evidence of anti-correlation so we concluded that the star-galaxy anti-correlation is only serious in star samples too close to the $W1$ limit. In this case the effect on our QSO samples will be small. But we shall show the star-galaxy correlation results alongside the quasar-galaxy correlation results so that the size of any possible systematic effect can be judged.

\subsection{CMB Lensing Data}

We use the 2018 release of the Planck lensing convergence baseline map, using the CMB-only minimum variance estimates of the lensing signal to scales of $l=4096$ \citep{Planck2018}, to perform cross-correlations with our galaxy, galaxy cluster, and LRG samples. Small angular scales correspond to a high $l$ value as $\theta\sim\frac{180\deg}{l}$. The Healpix $a_{lm}$ are first smoothed with a Gaussian filter with a FWHM of $15$ arcmin. We then convert this baseline Minimum Variance lensing map from the stored convergence spherical harmonics $a_{lm}$ to a Healpix map (as done by \citealt{Geach2019}) with nside$=2048$ and an $l_{max}=4096$. This then gives us a list of RA and Dec coordinates of the Healpix pixel centers. We apply the lensing mask provided by the \cite{Planck2018} to the CMB data and select two areas that overlap our $\sim4700$deg$^2$ QSO sample.

\subsection{Possible systematic effects}

Contamination of the QSO sample by stars or galaxies will show different effects on our cross-correlation results. Star contamination will dilute bright and faint cross-correlations by the fraction of stars in the QSO sample. However, the $grW1$ cut we make is very efficient at removing stars at the $g<22$ magnitude range of our QSO sample. So the main QSO contaminant is likely to be galaxies in the same redshift range as the $r<21$ galaxy sample and this will reduce galaxy QSO anti-correlation at faint magnitudes while increasing galaxy-QSO cross-correlation at bright QSO magnitudes. However, the restricted version of our quasar sample which we are using reduces  this contamination (see Section~\ref{sec:xcorr_QSO_sample}). We shall see that the level of agreement between the positive and negative cross-correlations seen at bright and faint QSO magnitudes with a lensing model can be taken as confirming this low level of galaxy contamination. \\

A similar argument applies to any dust obscuration associated with the foreground galaxy population, since this would increase the anti-correlation at faint QSO magnitudes while decreasing the positive signal at  bright magnitudes, producing disagreement with the lensing model. \cite{Menard2011} did find evidence for dust effects in the SDSS galaxy-QSO cross-correlations but they were highly sub-dominant with respect to the lensing effect. We tested limiting our QSO sample in the W1 band and compared the galaxy-QSO cross-correlations to those found in the $g$-limited QSO samples and again found little difference between the two, implying that lensing dominates our cross-correlation results. \\

The other major systematic was  the possible sky subtraction issue in W1,W2 in the vicinity of bright galaxies. This evidenced itself in a strong anti-correlation between bright galaxies and stars. However, the effect reduced when the star control sample was selected to have r-W1 colours more similar to the QSOs (see Section \ref{sec:xcorr_star_sample}) and we show these galaxy-star cross-correlations alongside the galaxy-QSO versions in Figs. \ref{fig:wgm_withHODmodels} and \ref{fig:wgm_withsp7HODmodels}, for comparison purposes.

\section{QSO - Galaxy Cluster Lensing}
\label{sec:QSOgalclust_lensing}

\subsection{Cross-Correlation Method}
\label{sec:CUTE}

We use the data samples described in Section~\ref{sec:xcorr_catalogues} to make a weak gravitational lensing analysis via a cross-correlation of background quasars and foreground galaxies and galaxy clusters. Following Limber's equation \citep{Limber1953}, we can express the 3-D correlation function (and power spectrum) as 2-D angular correlations. To calculate the angular cross-correlation, we need random data sets with the same input parameters as our  quasar +  galaxy/galaxy cluster samples. Therefore, we generate catalogues of uniformly distributed random  points covering the same area as our survey with typically $>10$ times as many sources as the observable data sets. These random catalogues are then also masked in the same manner as our data catalogues (see Fig.~\ref{fig:NGC_racut}).\\

We use the publicly available Correlation Utilities and Two-point Estimates (CUTE) code \citep{CUTE2012} to determine the angular cross-correlation of our samples. CUTE calculates the cross-correlation by using the normalized Landy-Szalay estimator for a two-point correlation function, defined as:

\begin{equation}
     \omega_{GQ}(\theta)=\frac{D_GD_Q-D_GR_Q-R_GD_Q-R_GR_Q}{R_GR_Q}, 
     \label{eq:LScorrelation}
\end{equation} \\

We check the output generated by the Landy-Szalay estimator by manually checking the $D_GD_Q$, $D_GR_Q$, $R_GD_Q$, and $R_GR_Q$ outputs which we need to calculate the angular cross-correlation. Here $D_GD_Q$ denotes the number of data-point pairs drawn from the galaxy sample and quasar sample with separation $\theta$. For $D_GR_Q$ the quasar sample is replaced with the sample of randomly distributed quasar points with the same angular selection function as the data. Similarly, for $R_GD_Q$ the galaxy sample is replaced with our random galaxy sample. The $R_GR_Q$ output is the number of data-point pairs drawn from the two random quasar and galaxy samples. 




To generate  error estimates from field-field variations, we divide the quasar and galaxy samples into  $N_s=8$ similarly sized $\approx600 deg^{2}$ regions, 4 in the NGC and 4 in the SGC. These fields are shown in Fig.~\ref{fig:NGC_racut}. Then we estimate the standard errors of the cross-correlation by using the field-field error:

\begin{equation}
     \sigma_{\Bar{\omega}(\theta)}=\frac{\sigma_{N_{s}-1}}{\sqrt{N_{s}}}=\sqrt{\frac{\sum(\omega_{i}(\theta)-\Bar{\omega_{i}}(\theta))^{2}}{N_{s}^{2}-N_{s}}}, 
     \label{eq:standard error}
\end{equation} \\

\noindent where the sum is over $i=1, N_s$.

\subsection{Quasar-Galaxy Cluster Lensing SIS Model}
\label{sec:QGalClustModel}


The lensing of the background objects depends on the mass profiles of the foreground objects. For galaxy clusters, we initially assume the simplest mass profile of a singular isothermal sphere (SIS). The deflection angle of sources by such foreground lenses  is given by:

\begin{equation}
     \alpha=\frac{4GM(<b)}{bc^2}=\frac{D_s}{D_{ls}}(\theta-\theta_q), 
     \label{eq:deflection_angle}
\end{equation} \\

\noindent (e.g. \citealt{Myers2003}) where $b$ is the impact parameter, $M(<b)$ is the mass contained within the radius of the lens, $D_s$ is the angular diameter distance from the observer to the source, $D_{ls}$ is the angular diameter distance from the source to the lens, $\theta$ is the angle from the observer's line of sight to the image, and $\theta_q$ is the angle from the observer's line of sight to the source quasar. \\ 


We see an increase in apparent brightness/magnitude of the background object as the surface brightness of the object is conserved, but spread across a larger surface area. Therefore the flux received from the object is increased. The magnification, A, of the object due to a foreground lens can be described as:


\begin{equation}
     A=\left|\frac{\theta}{\theta_q}\frac{d\theta}{d\theta_q}\right|
     \label{eq:amplification_factor}
\end{equation} \\


\noindent On the assumption of lensing by a SIS, the mass surface density is: 

\begin{equation}
     \Sigma_{SIS}=\frac{\sigma^2}{2Gr}  
     \label{eq:SIS}
\end{equation} \\

\noindent where $\sigma$ is the velocity dispersion of the SIS and the density goes as $\rho(r)=\frac{\sigma^2}{2\pi Gr^2}$. This can be integrated over a radius of $r=0$ to $r=b$ and combined with Eq.~\ref{eq:deflection_angle} to give the amplification  due to a SIS of a background source at radius $\theta$:

\begin{equation}
     A=\left|\frac{\theta}{\theta-4\pi(\frac{D_{ls}}{D_s})(\frac{\sigma}{c})^2}\right| 
     \label{eq:SIS_amplification}
\end{equation} 

\noindent This amplification factor can also be described as the ratio of the lensed flux and the unlensed flux \citep{Croom1997}. As the amplification affects the relative distribution of background and foreground objects, we can relate the angular cross-correlation to the amplification factor through:

\begin{equation}
     \omega(\theta)=A^{2.5\alpha-1}-1
     \label{eq:correlation_func}
\end{equation} 

\noindent where $\alpha$ is the slope of the cumulative source number count, $dlog(N)/dm$. Zero correlation is predicted at $\alpha=0.4$ with an anti-correlation at $\alpha<0.4$, and a positive correlation at $\alpha>0.4$.

In our model, we use the flat $\Lambda$CDM cosmology, with $\Omega_M=0.3$ and $\Omega_{\Lambda}=0.7$. We assume an average foreground galaxy sample and galaxy cluster redshift of $z=0.15$ and an average quasar sample redshift of $z=1.5$. This gives us an angular diameter distance of the quasar sample $D_S=1780$ Mpc and $D_{LS}=1235$ Mpc. We also use a lensing coefficient of $2.5\alpha-1=-0.37$ for the faint QSOs with $20<g<21$  and $2.5\alpha-1=0.95$ for the bright QSOs with $17<g<19$ taking these and other values from Table 1 of \cite{Scranton2005} for consistency with their assumptions. 

\subsection{Quasar-Galaxy Cluster Lensing NFW Model}
\label{sec:clusterNFWmodel}


Similar to  modelling the cluster lensing via SIS we next model the clusters using an \citep*{NFW1996} NFW model. We follow this route here because the HOD approach for clusters is less developed than for galaxies. Nevertheless, for computational convenience we use the Cosmology and HalO Model Python code (CHOMP), which is a halo modelling package written by \href{https://github.com/karenyyng/chomp/blob/master/README.txt}{Morrison, Scranton, and Schneider} to produce the projected, lensed NFW mass profile which in 3-D takes the form:

\begin{equation}
     \rho(r)=\frac{\rho_0}{(r/r_s)(1+r/r_s)^2}, 
     \label{eq:nfw}
\end{equation} \\

\noindent CHOMP also assumes that halo concentration is a function of halo mass with the functional form  $c(m)\approx9(m/m^*)^{-0.13}$ taken  from \cite{Bullock2001}. We then compute  these projected, lensed NFW mass profiles by simply 
isolating the NFW part of the 1-halo term produced by CHOMP. Full  details of the 1- and 2-halo terms and their projection and magnification as implemented in CHOMP are given by \cite{Jain2003} and will be further summarised in Section \ref{sec:HOD_models}.


\subsection{Quasar-Galaxy Cluster Cross-Correlation Results}

\begin{figure}
	\centering
	\includegraphics[width=\columnwidth]{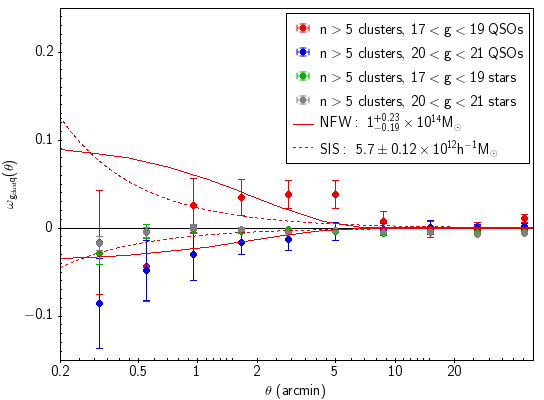} 
        \includegraphics[width=\columnwidth]{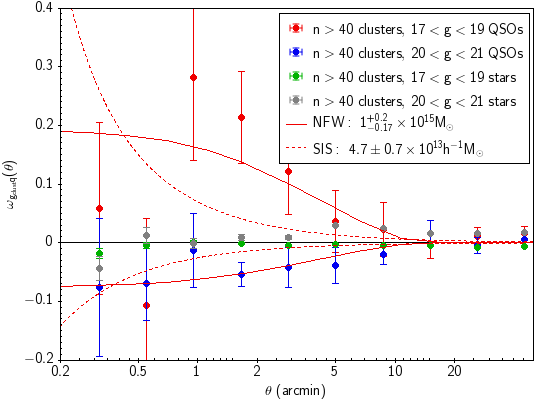} 
	\caption[Results of the cross-correlation of both our bright and faint quasar candidate catalogues in the $g-$band and the VST ATLAS galaxy cluster catalogue]{Results of the cross-correlation of both our bright and faint quasar candidate catalogues in the $g-$band and the VST ATLAS galaxy cluster catalogue for clusters comprised of n>5 and n>40 galaxies, using the CUTE code for angular cross-correlation. The SIS model here has a velocity dispersion of $270kms^{-1}$ and $460kms^{-1}$ and the HOD models are using a halo mass of $10^{14}$ solar masses for the the $n>5$ galaxy clusters, and a halo mass of $10^{15}$ solar masses for the $n>40$ galaxy clusters. The positive models are for the bright QSO-galaxy cluster cross correlation results and the negative models are for the faint QSO-gal clust cross correlation results.}
	\label{fig:qg_xcorr_n5n7n40}
\end{figure}

We perform the cross-correlation of our $n>5$ and $n>40$ galaxy cluster catalogues with our $17<g<19$ and $20<g<21$ quasar samples. We test the robustness of our detections by performing the cross-correlations with star samples in the same magnitude ranges. We can see in Fig. ~\ref{fig:qg_xcorr_n5n7n40} a clear anti-correlation with the faint, $20<g<21$, quasar samples for both the $n>5$ and $n>40$ galaxy clusters. The cross-correlations with the faint star samples show virtually zero correlation in comparison, making a strong argument for the reality of our detected cluster-quasar cross correlation signals at both bright and faint QSO magnitudes in Fig.~\ref{fig:qg_xcorr_n5n7n40}. 

We perform a $\chi^2$ test for both the SIS model and the NFW based HOD model on the cross-correlation results in order to  determine which model best describes our results in Figs. \ref{fig:qg_xcorr_n5n7n40}. To do this, we use the inverse variance weighted mean of the bright and faint QSO cross-correlation results for both the $n>5$ and $n>40$ galaxy cluster cases. 

For the SIS model, we find that the $n>5$ galaxy cluster - QSO cross-correlation has a best fit velocity dispersion of $\sigma=270^{+50}_{-65}$kms$^{-1}$ with a reduced $\chi^2$ of 1.8 and the $n>40$ galaxy cluster - QSO cross-correlation has a best fit velocity dispersion of $\sigma=460^{+60}_{-80}$kms$^{-1}$ with a reduced $\chi^2$ of 2.0. 
Using the $M=\frac{2\sigma^2r}{G}$ relation appropriate for an SIS model and taking $r=0.17$h$^{-1}$Mpc and $r=0.48$h$^{-1}$Mpc for $n>5$ and $n>40$ clusters respectively as empirically estimated from the cluster data themselves.\footnote{These cluster radii were empirically estimated from the cluster membership, assuming an $8\times$ overdensity on the sky, covering a circular area, and an average redshift of $z=0.2$ as suggested by \cite{Ansarinejad2023}.} These velocity dispersions correspond to masses of  $5.7\times10^{12}$h$^{-1}M_\odot$ for $n>5$ clusters and $4.7\times10^{13}$h$^{-1}M_\odot$ for $n>40$ (see Table \ref{tab:cluster_masses}).

For the NFW profiles, we similarly perform a $\chi^2$ fit to the $w_{cq}$, finding that the $n>5$ clusters are best fit by a mass of $10^{14\pm0.09}$h$^{-1}M_\odot$ with a reduced $\chi^2$ of 1.6 and the $n>40$ cluster cross-correlations are best fit by $10^{15\pm0.08}$h$^{-1}M_\odot$ with a reduced $\chi^2$ of 0.5. 

We conclude that the NFW is a better fit for the galaxy cluster-QSO cross-correlation as the SIS generally appears to be too steep at small scales, 
while the NFW is better able to fit the dampening of the signal at small scales. 
The implied NFW mass for $n>40$ clusters also is more in agreement with the mass estimates of \cite{Ansarinejad2023}, based on various calibrations of cluster membership, that gave a mean mass of our $n>40$ galaxy clusters of $4.3\pm2.7\times10^{14}$h$^{-1}$M$_\odot$ (see Table \ref{tab:cluster_masses}).

\begin{table}
\caption[Summary of results for galaxy cluster masses]{Summary of results for galaxy cluster masses. $w_{cq}$ denotes the cluster-QSO cross-correlation, shown for the SIS and NFW cases, and $w_{c\kappa}$ is the cluster-CMB cross-correlation. A2023 cluster masses are estimated by \cite{Ansarinejad2023}}
\label{tab:cluster_masses}
\begin{tabular}{llclc}
\hline
Method            &  n>5 Mass         & n>5                      &    n>40   Mass                 & n>40    \\         
                  & $(10^{13}$h$^{-1}M_{\odot})$& $\chi^2_{red}$ & $(10^{13}$h$^{-1}M_{\odot})$   &$\chi^2_{red}$ \\   
\hline             
$w_{cq}$ SIS      & $0.57\pm0.12$    &  1.8                      & $4.7\pm0.7$                    & 2.0  \\    
$w_{cq}$ NFW      &$10.0\pm2.1$      &  1.6                      & $100\pm20$                     & 0.5 \\ 
$w_{c\kappa}$ NFW & $3.2\pm0.7$      &  3.2                      & $32\pm12$                      & 1.1 \\   
A2023             & $23\pm8$         &  $-$                      & $43\pm27$                      & $-$ \\   
\hline
 \end{tabular}
\end{table}

\subsection{Galaxy Cluster - CMB Lensing Map Cross-Correlation}
\label{sec:cluster_cmb}

Galaxy cluster-QSO cross-correlation mainly probes the 1-halo term, whereas cross-correlation of the Planck CMB lensing convergence map with the galaxy clusters only constrains the 2-halo term due to the $\approx6'$ Planck resolution. Nevertheless, we can check if the NFW profiles found to fit our QSO-galaxy cluster cross-correlations give halo masses consistent with the CMB lensing method.  We model the CMB lensing by  foreground galaxy clusters using the  5-parameter  HOD methodology of \cite{Zheng2007}. We again employ the above CHOMP halo modelling package and here use it more conventionally, to make 1-halo + 2-halo predictions, with the latter dominant. We assume the following HOD parameters $\log(M_0)=\log(M_{min})$,  $\log (M_{1'})=\log(M_{min})+1.08$, $\sigma_M=0.4$ and $\alpha=0.7$ with masses in solar mass units assuming $h=0.7$. These parameters are used for values of $\log(M_{min})=12.0,12.5,13,13.5,14,14.5,15$ to probe a similar range of masses studied previously for both the SIS and NFW QSO lensing models. We assume a flat redshift distribution between $z=0.01$ and $z=0.36$ as an approximation for our cluster samples (see \cite{Ansarinejad2023}) and a flat redshift distribution is also assumed for the CMB between $z=1050$ and $z=1150$.

\begin{figure}
    \centering
    \includegraphics[width=\columnwidth]{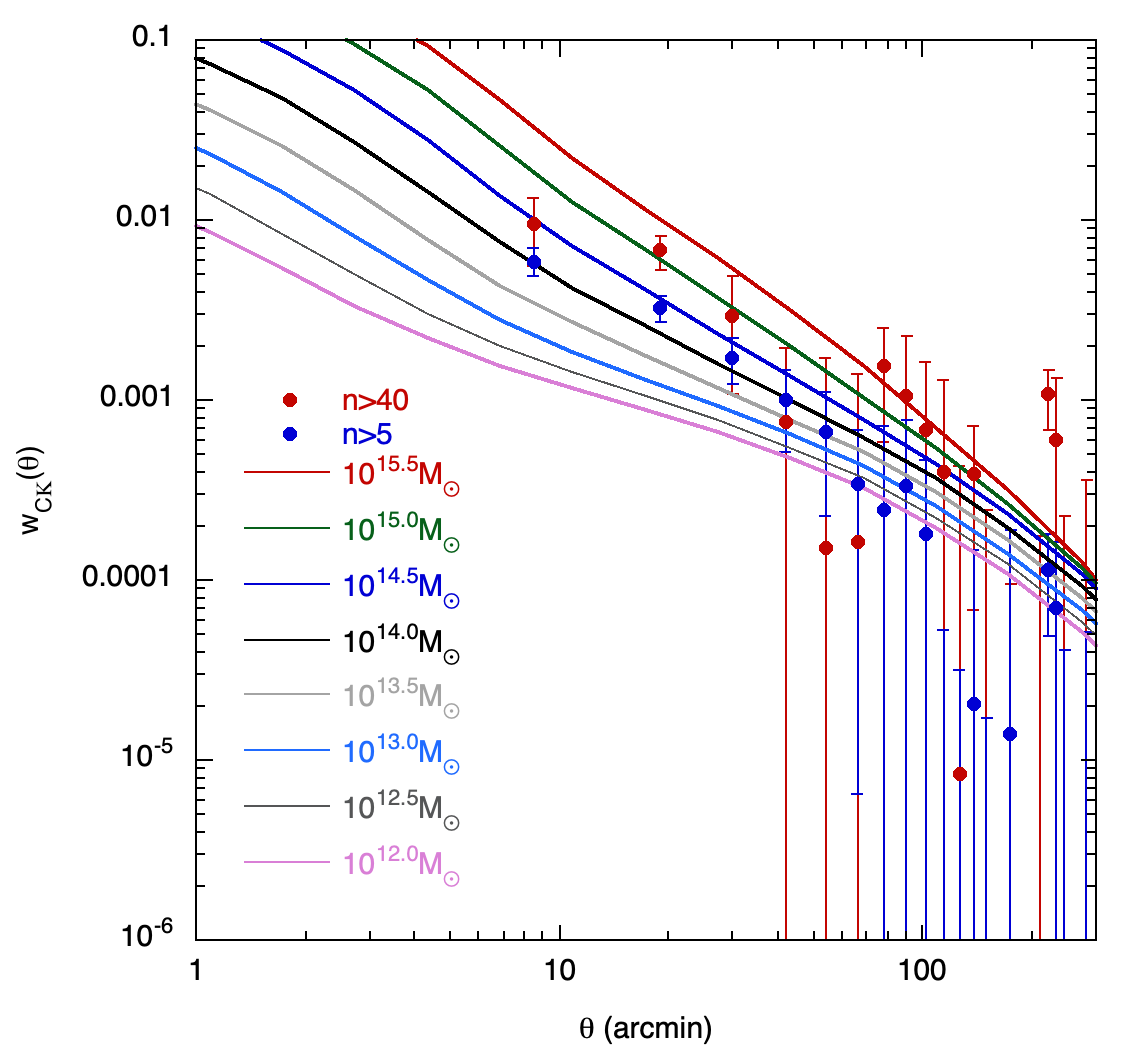} 
	\caption[Galaxy Cluster HOD models.]{$w_{c\kappa}$ cluster-CMB cross-correlation functions for $n>5$ and $n>40$ clusters compared to $w_{c\kappa}$ predicted by supplying CHOMP with the  \cite{Zheng2007}  HOD parameters of $\log(M_{min})=12.0,12.5,13,13.5,14,14.5,15$, $\log(M_0)=\log(M_{min})$,  $\log (M_{1'})=\log(M_{min})+1.08$, $\sigma_M=0.4$ and $\alpha=0.7$ with masses in solar mass units assuming $h=0.7$. The models were integrated over the redshift range $0.01<z<0.36$ and $\sigma_8=0.8$ was assumed throughout. }
	\label{fig:wCK_NFW_Mmp3}
\end{figure}

Shown in Fig. \ref{fig:wCK_NFW_Mmp3}, we see the result of  cross-correlating  the $n>5$ and $n>40$ galaxy clusters with the CMB lensing convergence map, along with the various HOD results. There is a potential smoothing at the smallest scales here due to the $6'$ resolution of the Planck CMB lensing convergence data. Therefore, the results in the bin at the smallest scale may be more systematically uncertain than indicated by the field-field error bars.

Performing a $\chi^2$ fit of the models to the data we find that the cross-correlation of the $n>5$ clusters with the CMB lensing convergence map is best fit by a HOD with $\log_{10}(M_{min})=13.5^{+0.09}_{-0.11}$ with a reduced $\chi^2$ of 3.2, which is not a good fit. For cross-correlation of $n>40$ clusters, we get a best fit model with $\log_{10}(M_{min})=14.5^{+0.14}_{-0.2}$ with a reduced $\chi^2$ of 1.1, with the corresponding NFW 1-halo term from QSO lensing giving  $\log(M_{h})=15$. In general, the cross-correlation of galaxy clusters with the Planck CMB lensing convergence map seem to agree with the NFW model results from QSO lensing in the previous section, although the $n>40$ fit has a slightly lower 2-halo mass than the NFW fit for the 1-halo term. We see a more significant departure in the halo mass predictions of the SIS model with the SIS masses being $\approx10\times$ smaller than the NFW masses, as summarised in Table \ref{tab:cluster_masses}.
The average masses of the $n>5$ and $n>40$ clusters as estimated by \cite{Ansarinejad2023} are also given in Table \ref{tab:cluster_masses}. We see that for $n>40$ clusters, our NFW lensing masses bracket the estimate of \cite{Ansarinejad2023} and so are in good agreement. For $n>5$ clusters the QSO and CMB  lensing masses are  a factor of $\approx2\times$ smaller than that of \cite{Ansarinejad2023} and so the agreement is less good here. 

We conclude  that for the richer, $n>40$, galaxy clusters, the NFW density profile fits significantly  better than the SIS profile at the small, 1-halo, scales probed by our QSO lensing results.  Generally the SIS profiles are too centrally peaked compared to the QSO lensing data. At larger scales, the CMB lensing results for these richer clusters also suggest that they are well-fitted by  a HOD model with a 2-halo term based on a $\Lambda$CDM cosmology. The estimated average mass for these richer clusters, assuming NFW/$\Lambda$CDM 1+2-halo terms, is in the range  $3\times10^{14} - 1\times10^{15}$h$^{-1}$M$_\odot$, in good agreement with mass estimates from \cite{Ansarinejad2023} and other authors. 

For the  less rich $n>5$ groups and clusters, the QSO lensing statistics are poorer and here both the 1-halo NFW and the SIS models provide acceptable fits to these data. The best-fitting NFW model implies a mass of $\approx1\times10^{14}$h$^{-1}$M$_\odot$ for this $n>5$ sample, a factor of $\approx2\times$ lower than the estimate of \cite{Ansarinejad2023} but in agreement  within the errors. At larger scales, the CMB lensing signal for this $n>5$ sample is strongly detected at a level almost as high as for the $n>40$ sample. However, in this case, a  HOD model based on a $\Lambda$CDM cosmology and where the minimum halo mass was allowed to vary in the range $1\times10^{12}<M_{\rm min}<3\times10^{15}$h$^{-1}$M$_\odot$  could not be found to fit the CMB lensing data when fitted over the full $\theta<300'$ range. The reason for this disagreement is currently unclear but will be further investigated in the work on galaxy lensing following in Sections \ref{sec:QSOgal_xcorr} and \ref{sec:HOD_models}.

\section{QSO-Galaxy Cross-Correlation}
\label{sec:QSOgal_xcorr}

We now turn to estimating foreground galaxy halo masses  via the lensing of background QSOs and the CMB, complemented by  constraints from the angular autocorrelation function of the same galaxies. For the galaxy-QSO cross-correlations, we shall first use a model where galaxies trace the mass to connect with the previous studies of, e.g. \cite{Myers2003}, before dropping this assumption and fitting HOD models (such as \citealt{Scranton2005, Jain2003, Zheng2007} etc).

\subsection{Quasar-Galaxy Cross-Correlation Model}
\label{sec:QGalModel}

We first use the \cite{WI1998} model, as outlined by \cite{Myers2005}, to describe the correlation between our quasar sample and foreground galaxies. Although \cite{Myers2005} uses a galaxy sample to $g<20.5$, we use a galaxy sample of $r<21$ in order to match the magnitude limit of the SDSS galaxy sample of \cite{Scranton2005}. This \cite{WI1998} (from here referred to as the WI model) bases predictions for $w_{gq}$ on  the auto-correlation, $w_{gg}$, of the galaxy sample and on the assumption that galaxies trace the mass. The lensing convergence $\kappa$ is defined as:

\begin{equation}
     \kappa=\frac{\Sigma(D_l,\theta)}{\Sigma_{cr}(D_l,D_s)}, 
     \label{eq:WI_kappa}
\end{equation} \\

\noindent where $D_l$ is again the angular diameter distance of the lens, $\Sigma(D_l,\theta)$ is the surface mass density of the lens, and $\Sigma_{cr}(D_l,D_s)$ is the critical mass surface density, defined in \cite{Myers2005} as $\Sigma_{cr}(D_l,D_s)=\frac{c^2}{4\pi G}\frac{D_s}{D_lD_{ls}}$. \\

We can estimate the effective convergence using the relation :

\begin{equation}
     \kappa_{eff}(\theta)=\frac{3H_0^2c}{8\pi G}\Omega_m(\delta_G-1)\int^{z_{max}}_0\frac{(1+z)^3\frac{dt}{dz}dz}{\Sigma_{cr}(z,z_s)}, 
     \label{eq:k_eff}
\end{equation} \\

\noindent (see \citealt{Myers2005, WI1998}). Here, we take $z=1.5$ as the median redshift of our quasar sample and the galaxy sample peaks at $\sim0.2$, so we integrate to a redshift of $z_{max}=0.3$ where the distribution drops to $\sim20\%$.   
From this calculation, we find $\Bar{\kappa}=0.025$. The quasar-galaxy cross-correlation can then be modelled using the $\omega_{gg}$ and a Taylor expansion of Eq.~\ref{eq:correlation_func}. Therefore we predict the galaxy-quasar cross-correlation using:

\begin{equation}
     \omega_{gq}(\theta)=(2.5\alpha-1)\frac{2\Bar{\kappa}}{b}\omega_{gg}(\theta), 
     \label{eq:w_gg}
\end{equation} \\

\noindent where $\frac{\Bar{\kappa}}{b}=\frac{\kappa_{eff}(\theta)}{(\delta_G-1)}$. Here $b$ represents the linear galaxy bias $b=\langle\delta_G-1\rangle/\langle\delta_M-1\rangle$. The r.m.s. galaxy fluctuation $\langle\delta_G\rangle$ will be estimated via $\omega{gg}$, here represented by a power law fit to our galaxy sample acf which  gives $\omega_{gg}=0.142\theta^{-0.70}$ in the range $\theta<120'$, as shown in Sec. \ref{sec:HOD_models}.

In passing, we note the excellent agreement of the ATLAS $17<r<21$ galaxy $w_{gg}$  with the equivalent SDSS  $w_{gg}$ of  \cite{Wang2013} also shown  in Fig. \ref{fig:wgm_withHODmodels} (a). Given this SDSS-ATLAS acf agreement  extends to $\theta=8$deg. or $r_{com}\approx90$h$^{-1}$Mpc at the average galaxy redshift of $z\approx0.22$, this represents a strong argument for the accuracy of these two independent results and also for the reliability of their parent datasets.



\begin{figure}
\centering
	\includegraphics[width=\columnwidth]{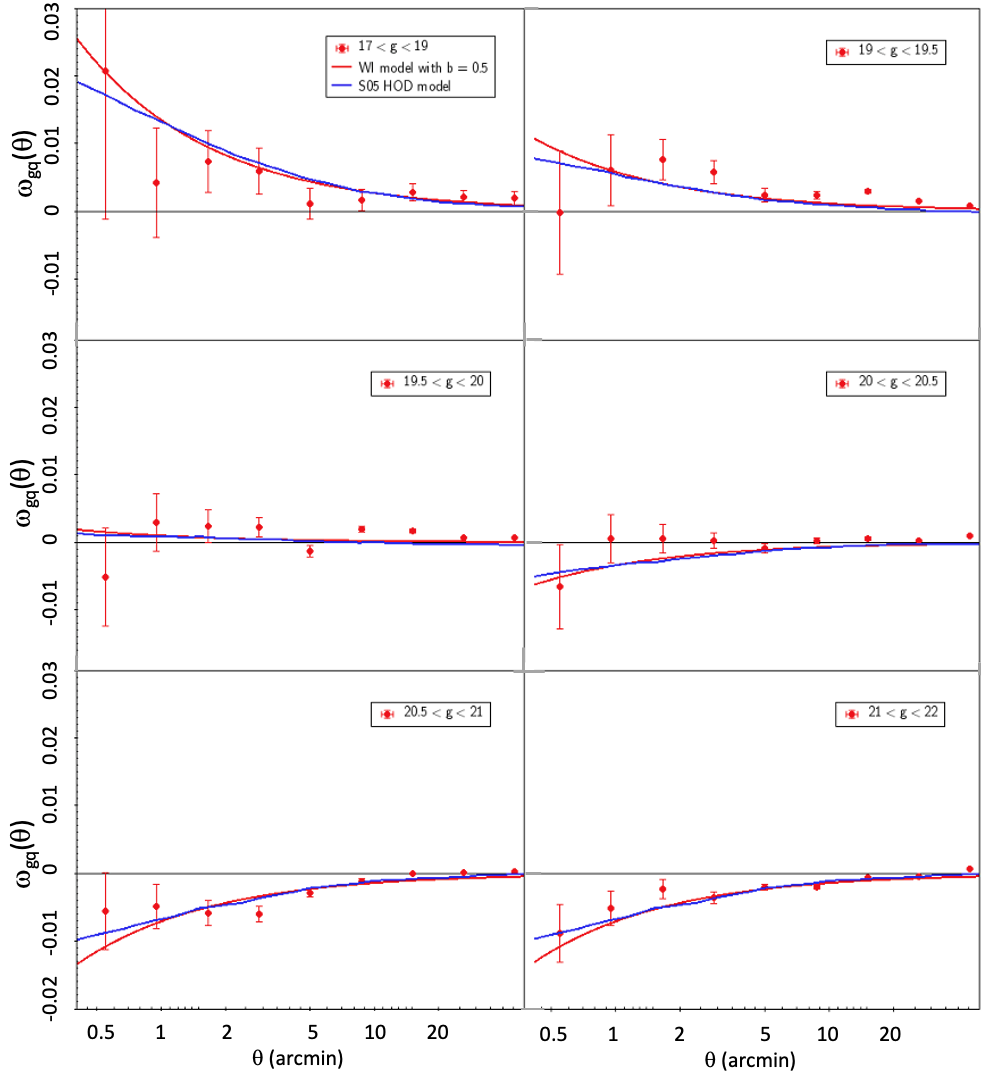} 
	\caption{Cross-correlations of our quasar candidate catalogue at $17<g<19$, $19<g<19.5$, $19.5<z<20$, $20.5<g<21$, and $21<g<22$ and our VST ATLAS galaxy catalogue at $r<21$, using the CUTE code for angular cross-correlation across the full sky. We also add the \protect\cite{Scranton2005} HOD model for each of the quasar $g-$band magnitude bins. A bias value of $b=0.5$ is consistently assumed for our WI model in red. The $\langle 2.5\alpha-1 \rangle$ values for each QSO magnitude range for both our model and the \protect\cite{Scranton2005} model are as follows: 0.95 for QSOs in the $17<g<19$ range, 0.41 for $19<g<19.5$, 0.07 for $19.5<g<20$, -0.24 for $20<g<20.5$, and -0.5 for $20.5<g<21$. We also assume this -0.5 value for the $21<g<22$ range.} 
	\label{fig:wgq_withmodels}
\end{figure}

\begin{figure}
    \centering
	\includegraphics[width=\columnwidth]{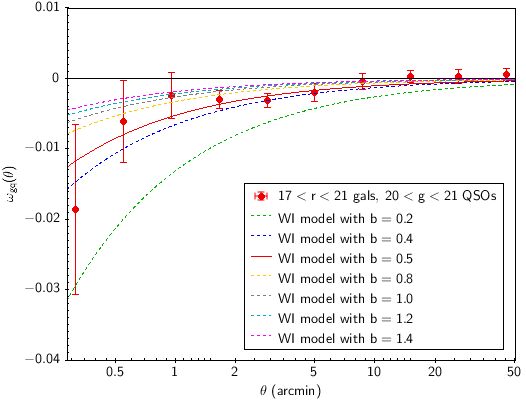} 
	\caption{Our $\omega_{gg}=0.142\theta^{-0.70}$ model fit, with $\Bar{\kappa}=0.025$ and $\langle\alpha-1 \rangle$=-0.37, with bias values of $b=0.2$, $b=0.4$, $b=0.5$, $b=0.8$, $b=1.0$, $b=1.2$, and $b=1.4$ for our cross-correlation at $20<g<21$.}
	\label{fig:crosscorr_withbiasmodels}
\end{figure}

\subsection{Quasar-galaxy cross-correlation results}

The results of cross-correlating  our  ATLAS QSO  catalogue in various  magnitude ranges with our $17<r<21$ mag galaxy catalogue is shown in Fig. \ref{fig:wgq_withmodels}. Also shown is the HOD model from the SDSS results of \cite{Scranton2005} in blue, and the WI model described in the previous section is shown in red. At angular scales of $\theta<5'$, we see a negative cross-correlation between ATLAS quasars and foreground galaxies at quasar g-band magnitudes of $g>20$ whereas at brighter QSO limits we see a positive correlation. These are the same trends as seen by \cite{Scranton2005} and by \cite{Myers2003,Myers2005} previously and they are as expected on the basic theoretical lensing model described in Section \ref{sec:QGalModel}.

\begin{figure}
    \centering
    \includegraphics[width=0.9\columnwidth]{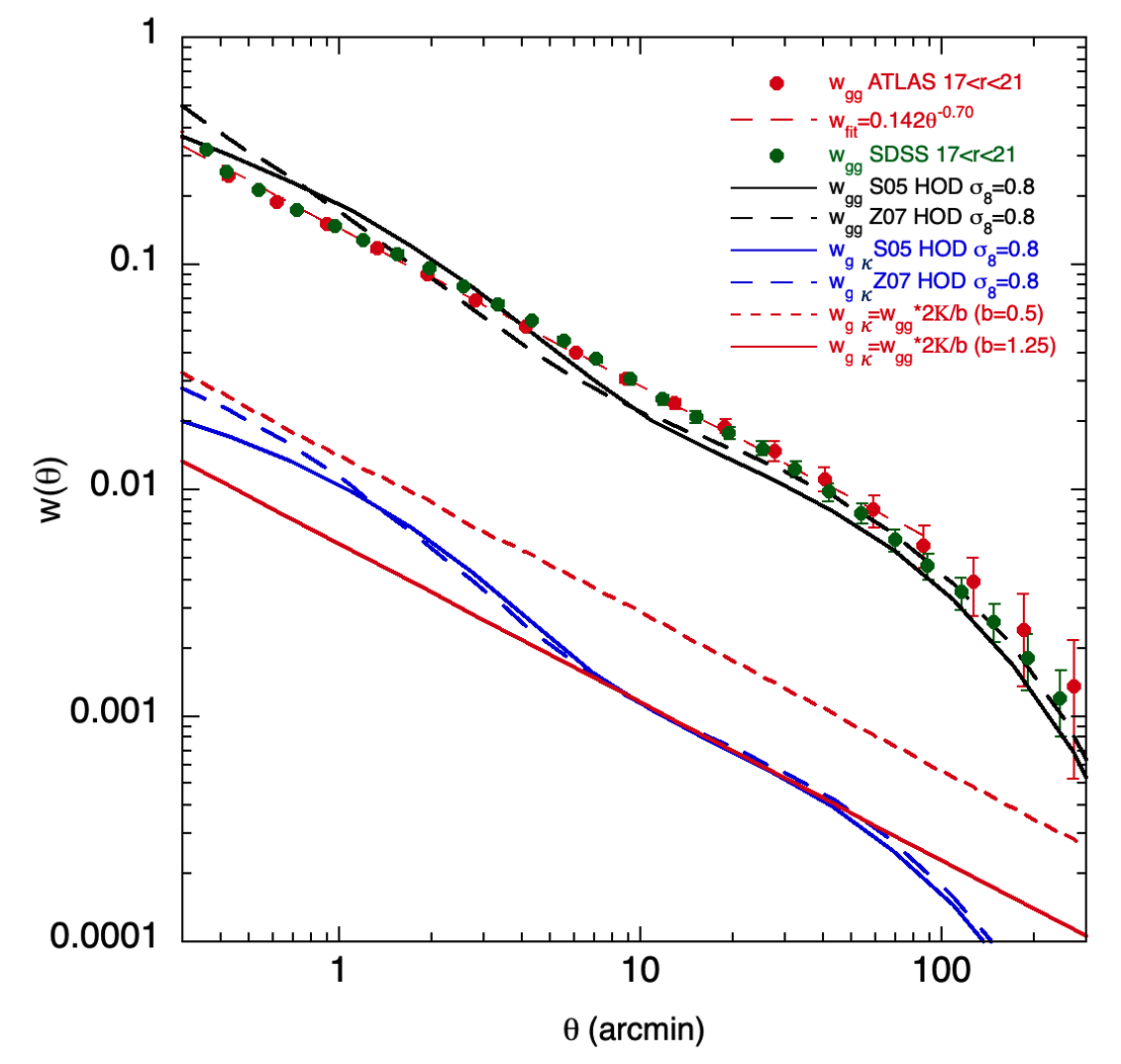}
    \includegraphics[width=0.9\columnwidth]{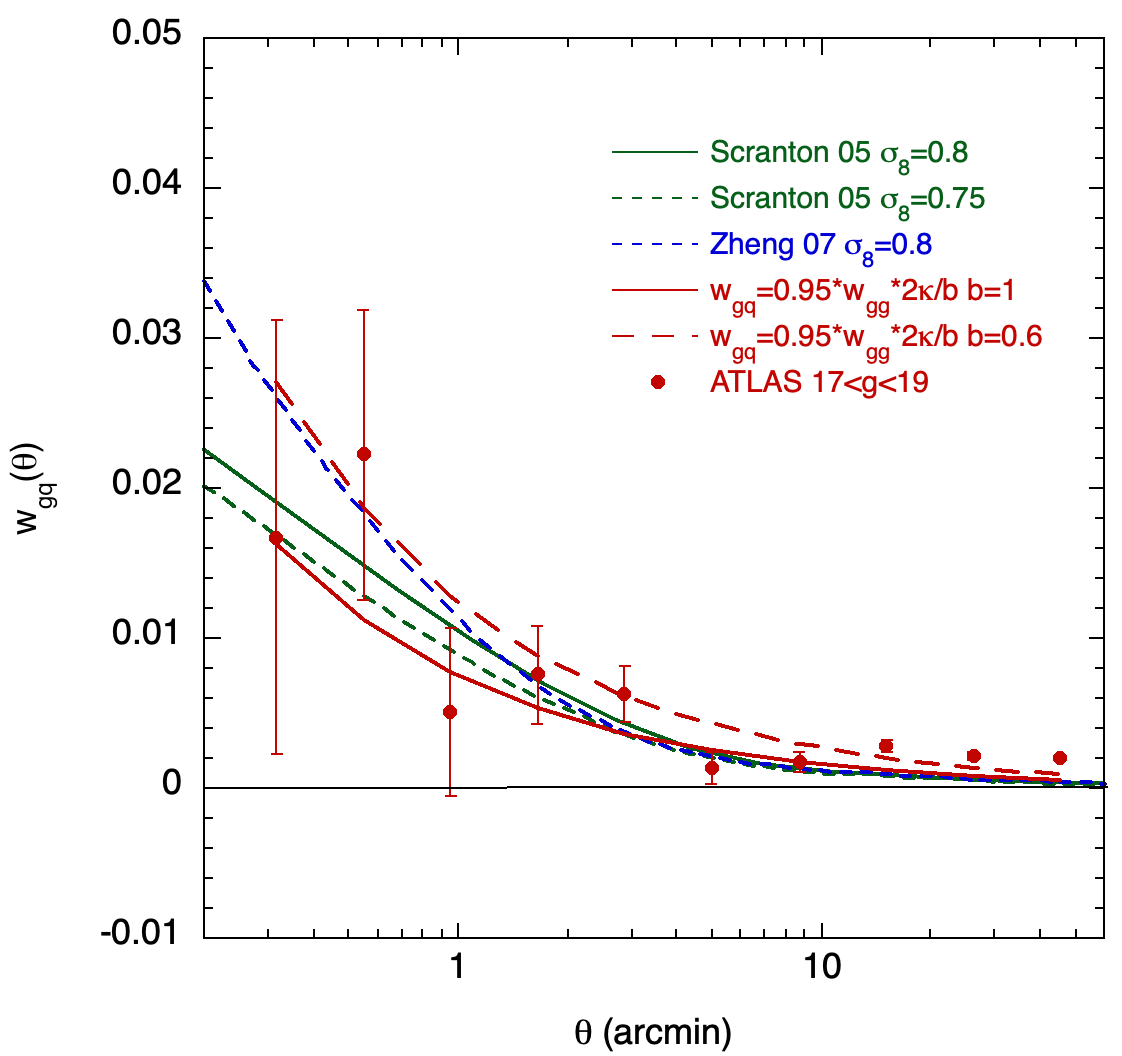}
 	\includegraphics[width=0.9\columnwidth]{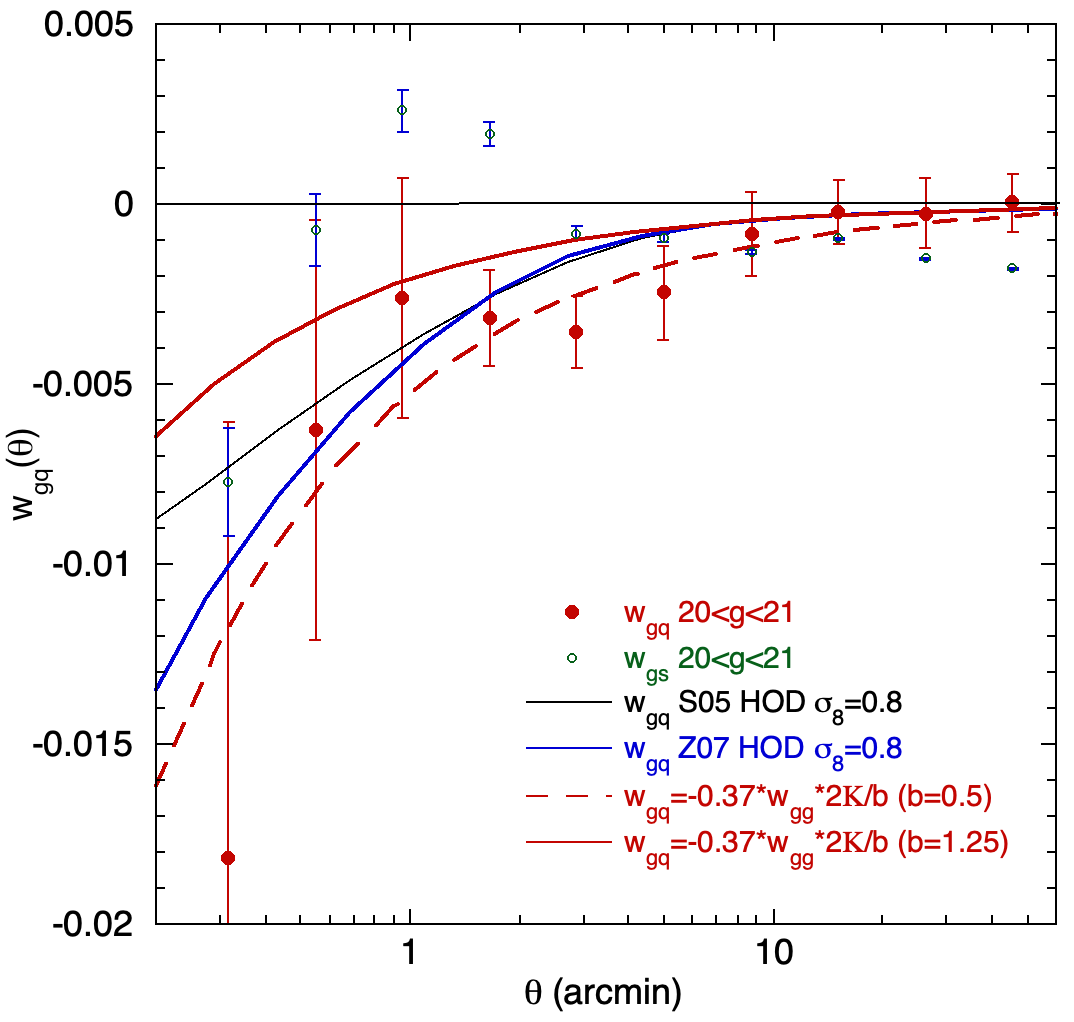}    \vspace{-0.4cm}
	\caption[HOD models.]{(a) $w_{gg}$ and $w_{g\kappa}$ auto- and cross-correlation functions predicted by the HOD models of \cite{Scranton2005} 
  and Zheng, Coil \& Zehavi (2007) (with $M_r<-20$).  Both models assume $h=0.7$ and $\sigma_8=0.8$. The $w_{gg}$ model fitted for the $17<r<21$ galaxies is $w_{gg}(\theta)=0.142\theta^{-0.70}$ (red, long dashes). (b) The cross-correlation function, $w_{gq}(\theta)$, for $17<g<19$ QSO candidates and $17<r<21$ galaxies, compared to the two HOD models and the two models of \cite{WI1998} with $b=1.25$ and $b=0.75$. (c) The same as (b) for the $20<g<21$ limited QSO case.}
	\label{fig:wgm_withHODmodels}
\end{figure}

To ease model comparisons between  \cite{Scranton2005} and ourselves, we use the values for $\langle\alpha_{S05}-1\rangle$=$\langle2.5\alpha-1\rangle$\footnote{Note that $\alpha_{S05}$ refers to a flux limited power law QSO number count, $N(>f)$ in the notation of \cite{Scranton2005}, whereas in our notation $\alpha$ refers to a magnitude limited power law number count, $N(<m)$.} listed in Table 1 and Fig. 2 of \cite{Scranton2005}. Then, using our $w_{gg}=0.142\theta^{-0.70}$ fit, with $\Bar{\kappa}=0.025$ and $(2.5\alpha-1)$=-0.37, we see from Fig. \ref{fig:crosscorr_withbiasmodels} that the best fit for the galaxy bias is  $b=0.5^{+0.13}_{-0.09}$ for the ATLAS cross-correlation at $20<g<21$.  Here, the fit  based  on   9(5) points in the range $\theta<30'(4')$, yields  low reduced $\chi^2=0.4(0.6)$. Also in this fit  the covariance between the $\omega_{gq}$ points is ignored since it is usually sub-dominant, due to the low space density of quasars \citep{BFS1988}. As previously noted, these cross-correlation amplitudes are high as measured  by the simple WI model since $b=0.5$ corresponds to $\sigma_8\approx2$\footnote{ If we assume $\sigma_{gg,8}\approx1$ then $b=1.25$ corresponds to $\sigma_8=\sigma_{gg,8}/b\approx0.8$ whereas $b=0.5$ corresponds to $\sigma_8\approx2$.} when the usual range is $0.7<\sigma_8<0.8$ (e.g. \citealt{Planck2020, Heymans2021}) i.e. $1.25<b<1.4$.\footnote{We note  that assuming $\Omega_m=1$ in eq \ref{eq:k_eff} would also increase the cross-correlation amplitude and imply a fitted bias value of $b\approx1.7$. Although this value is close to the expected $b=2$ for this cosmology, this $\Omega_m=1$ model is excluded by CMB + $H_0$ constraints and so we restrict our attention here to the standard cosmological model with $\Omega_m=0.3$.} Certainly, the $b=1.25$ (i.e. $\sigma_8\approx0.8$) model appears to give a poor fit in  Fig. \ref{fig:wgm_withHODmodels}(c), with the 5 points at $0.'5<\theta<8'$ giving a reduced $\chi^2=2.60$, rejecting the model at the 5\% significance level. However, dropping the assumption that galaxies trace mass may mean that models can be found that are more consistent with $\Lambda$CDM. 

So as previously suggested by \cite{Mountrichas2009}, we first conclude that there is little disagreement in terms of the observed data between SDSS and our ATLAS results and that the main disagreement is between these two models. We further conclude that the \cite{WI1998} assumption that galaxies trace the mass is unlikely to be correct, given that would imply $b=0.5$ i.e. $\sigma_8=2$ in contradiction with all observed CMB power spectra. So models that drop this assumption, like the  S05 HOD model, are likely to be required. However, the S05 HOD model may still underestimate the lensing signal, particularly at small $\theta<0.'5$ scales. So in Section \ref{sec:HOD_models}  we shall look for a HOD model  that improves the $w_{gq}$  fit while also simultaneously fitting the $w_{gg}$ of our $17<r<21$ galaxy sample.

\begin{figure}
    \centering
 	\includegraphics[width=\columnwidth]{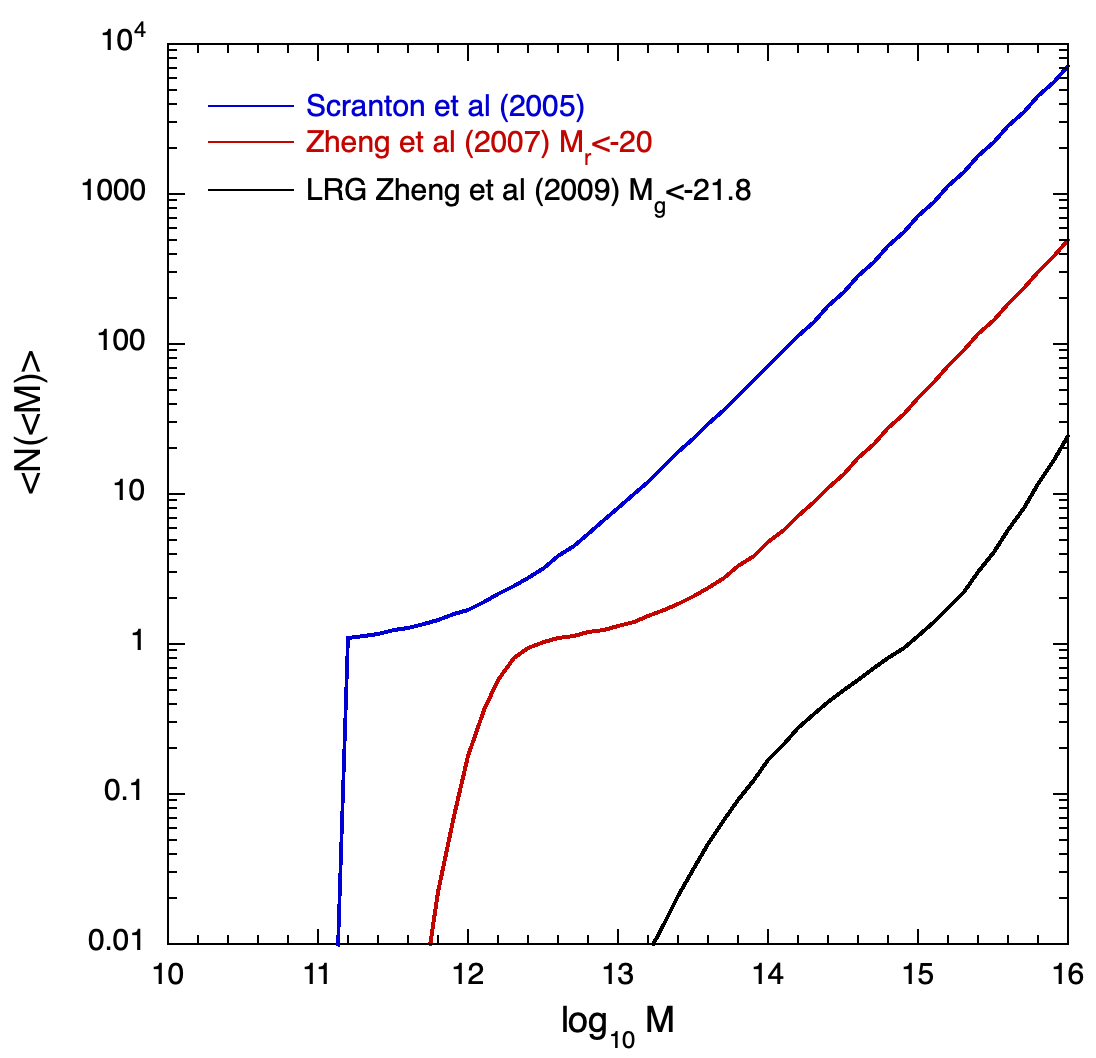}
	\caption[HOD models.]{ HOD models of Scranton et al (2005) (with ${\rm log(M_{min})}=11.15(11.0)$, $\sigma_M=0.01$, ${\rm log(M_0)}=0.0$,  ${\rm log(M_1')}=12.15(12.0)$, $\alpha=1.0$) and Zheng et al (2007) $M_r<-20$ model (with ${\rm log(M_{min})}=12.17(12.02)$, $\sigma_M=0.26$, ${\rm log(M_0)}=11.53(11.38)$,  ${\rm log(M_1')}=13.46(13.31)$, $\alpha=1.06$). The LRG model is from Zheng et al (2009) (with ${\rm log(M_{min})}=14.45(14.30)$, $\sigma_M=0.71$, ${\rm log(M_0)}=12.64(12.49)$,  $\log(M_1')=15.10(14.95)$, $\alpha=1.35$).  All masses assume $h=0.7$ ($h=1$).}
	\label{fig:HOD_NM_models}
\end{figure}

\section{HOD models via quasar-galaxy lensing and galaxy-galaxy clustering}
\label{sec:HOD_models}

\subsection {Modelling galaxy-galaxy angular correlations}

We now make a further check of the \cite{Scranton2005} HOD model using their publicly available code from the CHOMP GitHub site written by \href{https://github.com/karenyyng/chomp/blob/master/README.txt}{Morrison, Scranton, and Schneider}. The code follows \cite{Jain2003} in making predictions for both the angular auto-correlation function $w_{gg}$ and the galaxy-mass cross-correlation function $w_{g\kappa}$ based on a mass power-spectrum, $P(k)$, and a HOD, with the average number of galaxies per halo of mass $M$ being denoted by $<N(M)>$. 

First, we have assumed the simple  HOD model $<N(M)>=1+(M/10^{12.15})^{1.0}$  for $M>10^{11.15}M_\odot$ (with $h=0.7$) used  by \cite{Scranton2005} and  we use this to predict $w_{gg}$  for the $17<r<21$ galaxy sample used here (see Fig. \ref{fig:wgm_withHODmodels}a). We note in passing that \cite{Scranton2005} did not compare their observed and predicted $w_{gg}$. We found that this model with $\sigma_8=0.8$ over-predicted $w_{gg}$ at $\theta<5'$ and under-predicted it at larger, $\theta>5'$, scales. This under-prediction of the 2-halo term relative to the 1-halo term seems a common characteristic of HOD models. 
Essentially, the observed $w_{gg}$ seems to show a more exact power-law behaviour than the HOD models. \cite{MV2021} and references therein suggest that halo models generally underpredict the $\Lambda$CDM power-spectrum in the region between the 1- and 2-halo terms. Indeed, \cite{Peebles1974, Peebles1980} expressed doubts as to whether a preferred (halo) scale could ever be produced by the smooth $1/r^2$ power-law behaviour of Newtonian gravity. 

In searching for an improved HOD model, we then considered the HOD  recommended for SDSS galaxies with $M_r<-20$ by \cite{ZhengCoilZehavi2007} as an alternative to the simple S05 HOD. The parameters of this model are given in the caption of Fig. \ref{fig:HOD_NM_models}. This model produces slightly improved agreement with the ATLAS $w_{gg}$ at both small and large scales. We also considered the range of HOD models fitted to SDSS semi-projected correlation functions $w_p(\sigma)$ by \cite{Zehavi2011} (see their Fig. 10 and Table 3) corresponding to galaxies with absolute magnitudes from $M_r<-18.0$ to $M_r<-22.0$ but no better fit to our $17<r<21$ $w_{gg}$ was found. 

In more general searches within  the 5-parameter HOD scheme of \cite{ZhengCoilZehavi2007}, we still found it difficult to improve on the above SDSS $M_r<-20$ HOD as a description  of the ATLAS $w_{gg}$. Given the excellent agreement of the ATLAS $w_{gg}$ and the SDSS $w_{gg}$ of \cite{Wang2013}, also shown in Fig. \ref{fig:wgm_withHODmodels}a, we have no reason to believe that this  HOD fitting issue stems from the ATLAS data. So, bearing in mind  these  residuals at small and large scales, we shall consider the above two HOD models as reasonable fits and proceed to test them further using our weak lensing analyses.\footnote{Fitting a -0.8 power law to our $w_{gg}$ at $\theta<60'$ and then applying Limber's formula gives a 3-D correlation function scale-length of $r_0=5$h$^{-1}$ Mpc.} 

\subsection {HOD modelling from galaxy-quasar lensing}

We then continue to follow the method of \cite{Jain2003} to predict the $w_{g\kappa}$ cross-correlations, first assuming the \cite{Scranton2005} HOD. Having multiplied the model $w_{g\kappa}$'s in Fig. \ref{fig:wgm_withHODmodels}(a) by $(2.5\alpha-1)=0.95,-0.37$ for the bright $17<g<19$ and faint $20<g<21$ QSO samples, we compare the \cite{Scranton2005} and \cite{Zheng2007} HOD predictions to our $w_{gq}$ results in Figs. \ref{fig:wgm_withHODmodels}(b, c). In turn, we compare these to the $w_{gm}=w_{gg}\times2\Bar{\kappa}/b$ \cite{WI1998} models with $b=0.5$ and $b=1.25$. The \cite{ZhengCoilZehavi2007} HOD model seems to give a better fit than the \cite{Scranton2005} model in Figs. \ref{fig:wgm_withHODmodels}(b, c). with both models fitting these data better than the standard $b=1.25$ ($\sigma_8=0.8$) \cite{WI1998} model. Indeed, in Figs.  \ref{fig:wgm_withHODmodels} (b),(c) we  see that the HOD model of \cite{ZhengCoilZehavi2007} gives almost as good a fit as the best fit, $b=0.5$, \cite{WI1998} model. However, the errors are still large  in Figs. \ref{fig:wgm_withHODmodels}(b, c) and we remain wary about the size of the small-scale ($\theta<0.'5$) anti-correlation of the stellar control sample in Fig. \ref{fig:wgm_withHODmodels}(c). Another issue is that looking back at Fig. \ref{fig:wgm_withHODmodels}(a) we note that, at $\theta>1'$,  the predicted $w_{g\kappa}$ for the two HOD models with $\sigma_8=0.8$ lies significantly below the best fit, $b=0.5$ (or $\sigma_8=2$), \cite{WI1998} model implying that both sets of models cannot fit the data equally well on these larger scales.  This motivates a more detailed  study of the 1-halo term using LRGs in Section \ref{sec:lrg_hod_model} below, while a further test of the 2-halo fit of the HOD models is available from the CMB lensing test in Section \ref{sec:Gal_CMB_HOD} below. However, our main conclusion at this point is that we confirm that HOD models can be found that simultaneously give reasonable fits to $w_{gg}$ and   $w_{gq}$ at small scales and that these fit $w_{gq}$ significantly better than simpler models that assume galaxies trace the mass with bias in the standard $\Lambda$CDM  $b\approx1.2-1.4$ (or $\sigma_8\approx0.7-0.8$) range.

\subsection{Further Galaxy-CMB lensing test of HOD models}
\label{sec:Gal_CMB_HOD}

\begin{figure}
    \centering
 	\includegraphics[width=0.9\columnwidth]{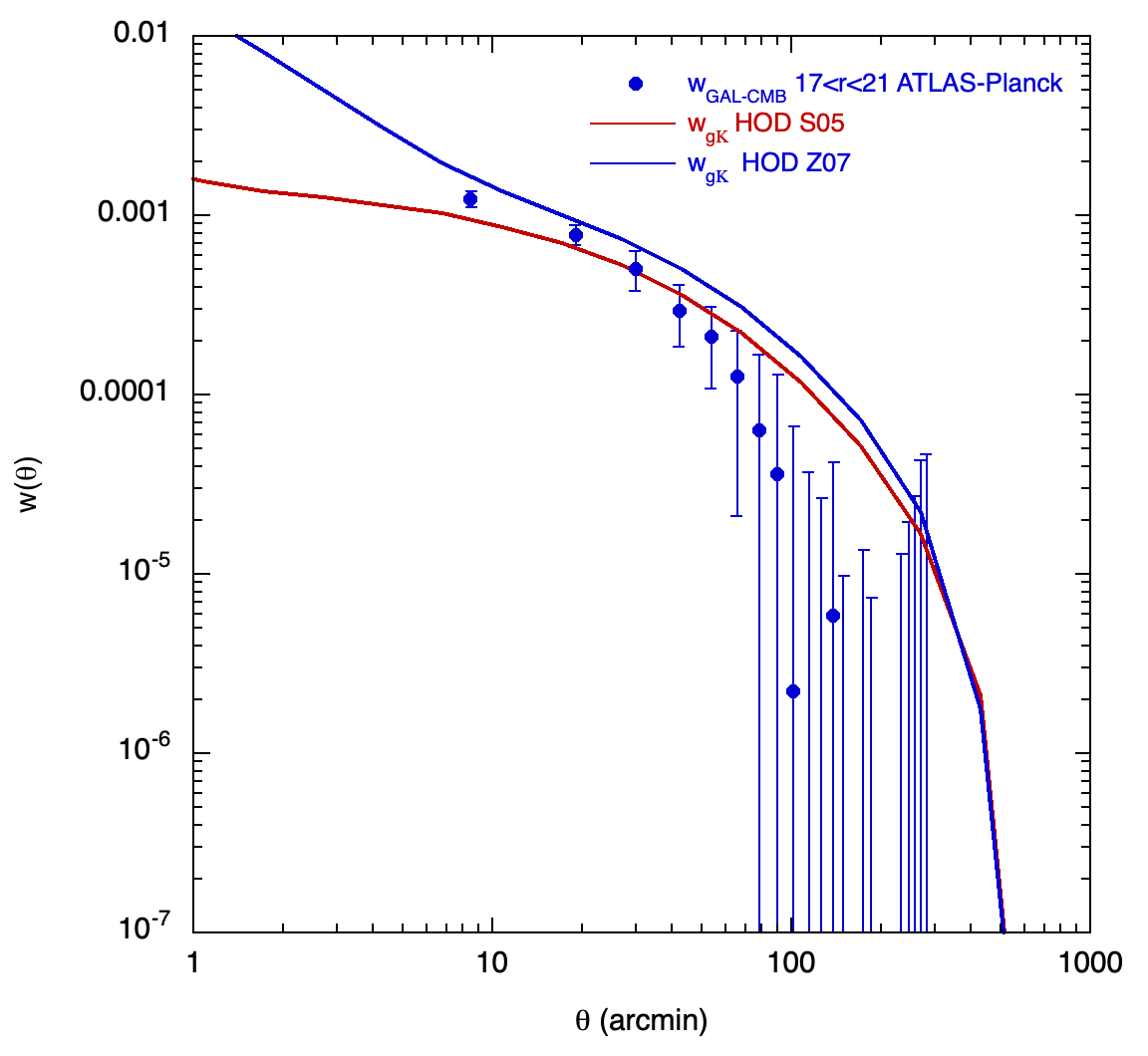} 
	\caption[HOD models.]{ $w_{g-CMB}$  cross-correlation function for $17<r<21$ galaxies and the Planck (2018) Lensing Map with field-field errors, compared to the HOD models of \cite{Scranton2005} and 
\cite{ZhengCoilZehavi2007} (with $M_r<-20$).  Both models assume $h=0.7$ and $\sigma_8=0.8$. The first bin centre at 6$'$ corresponds to $\approx1$ h$^{-1}$Mpc at the galaxy mean $z$ of $z=0.15$.}
	\label{fig:w_gcmb_withHODmodels}
\end{figure}

In Fig. \ref{fig:w_gcmb_withHODmodels} we show the $17<r<21$ galaxy - Planck CMB Lensing Map cross-correlation function compared to the predictions of the HOD models of \cite{Scranton2005} and \cite{ZhengCoilZehavi2007}. Here we see that the data is in reasonable agreement with  the \cite{Scranton2005} model at all scales (with a reduced $\chi^2$ of 2.67) and fits particularly well in the range $10'<\theta<60'$ with a reduced $\chi^2$ of 1.15, whereas the \cite{ZhengCoilZehavi2007} model appears to over-predict the data at all scales (with a reduced $\chi^2>10$), despite its good fit to $w_{gg}$ at $\theta>20'$. We also note that the scales probed with the {\it Planck} map are mostly at the scales of the 2-halo term with $r_{com}\approx1$h$^{-1}$ Mpc corresponding to $\theta\approx8'$ at the average galaxy redshift of $z=0.15$. So  CMB lensing  at {\it Planck} resolution  is clearly the test of choice for the 2-halo term while the galaxy QSO cross-correlation function in Fig. \ref{fig:wgm_withHODmodels} (b),(c), with its scale extending down to $\approx1'$, provides a better test of the 1 halo-term. Here we have seen that both S05 and Z07 models give reasonable fits to $w_{gq}$ but the S05 HOD fits the CMB lensing data better than the Z07 HOD galaxy  at larger scales, despite both HOD models fitting the $w_{gg}$ equally well in this range dominated by the 2-halo term. But higher signal-noise data for QSO lensing and  higher resolution data for CMB lensing  should give further interesting tests of both the 1- and 2-halo terms of these galaxy halo occupation  models independently over the full range of scales.

\section{LRG HOD modelling}
\label{sec:lrg_hod_model}

\begin{figure}
    \centering
  	\includegraphics[width=0.9\columnwidth]{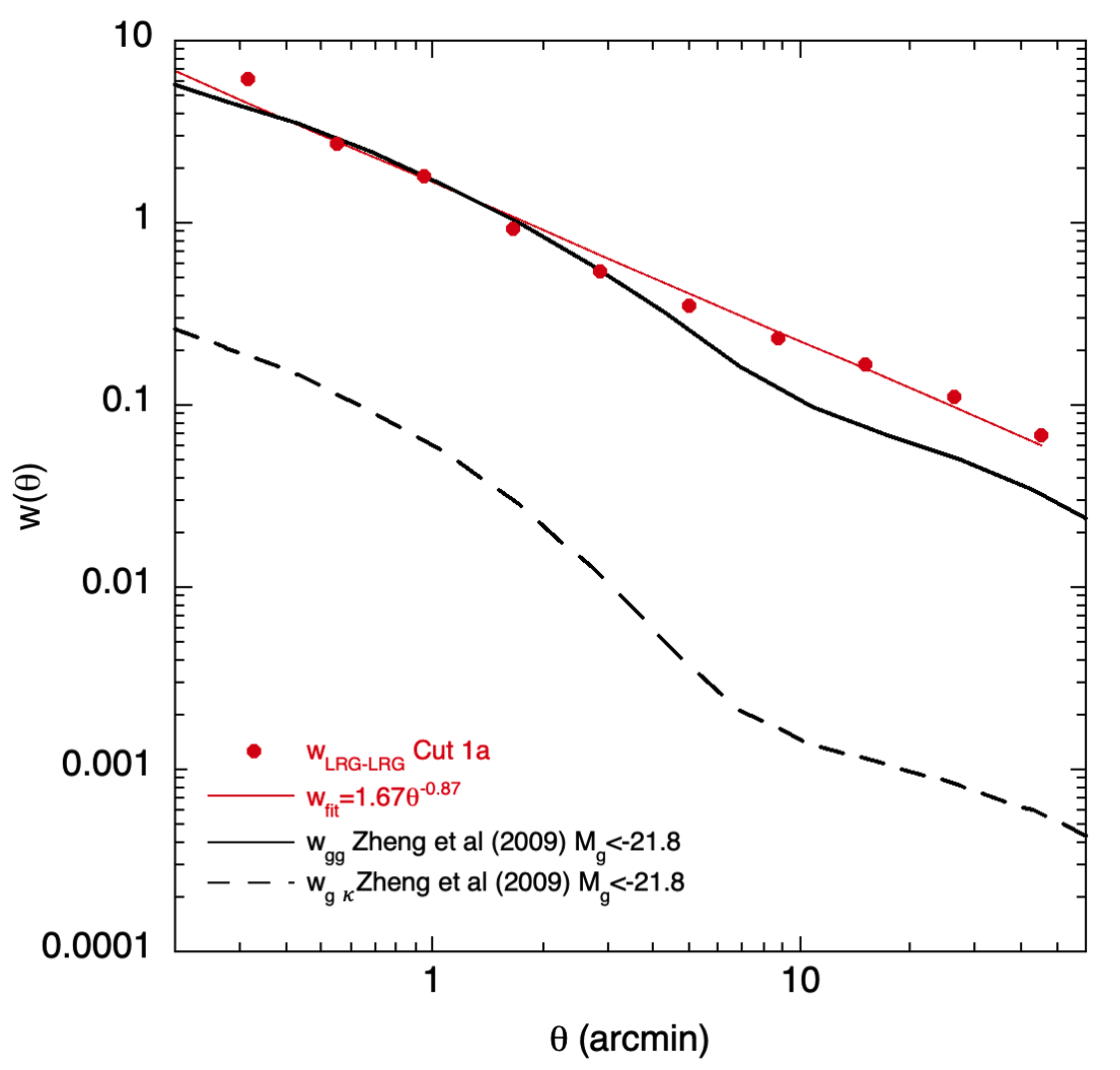}
  	\includegraphics[width=0.9\columnwidth]{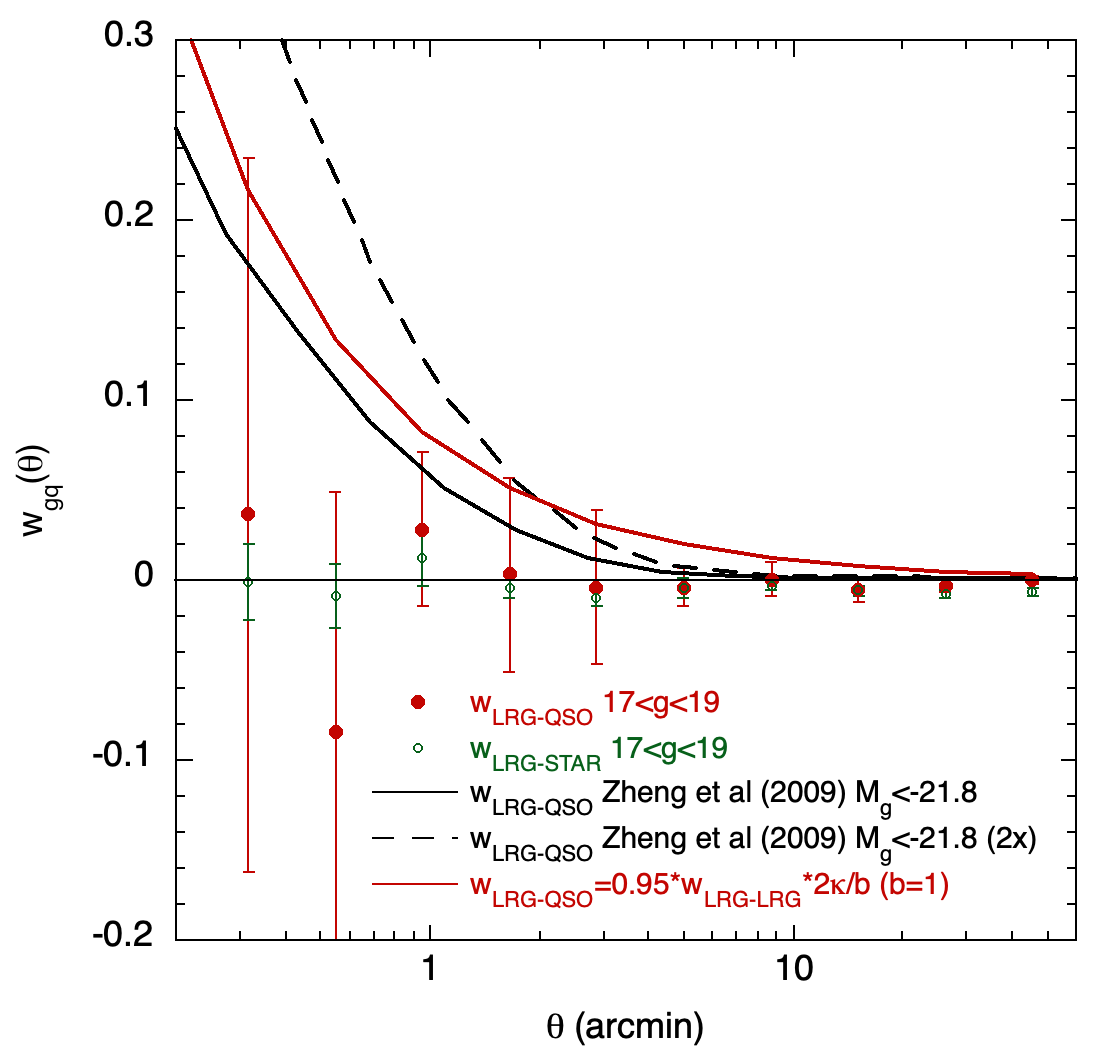}
  	\includegraphics[width=0.9\columnwidth]{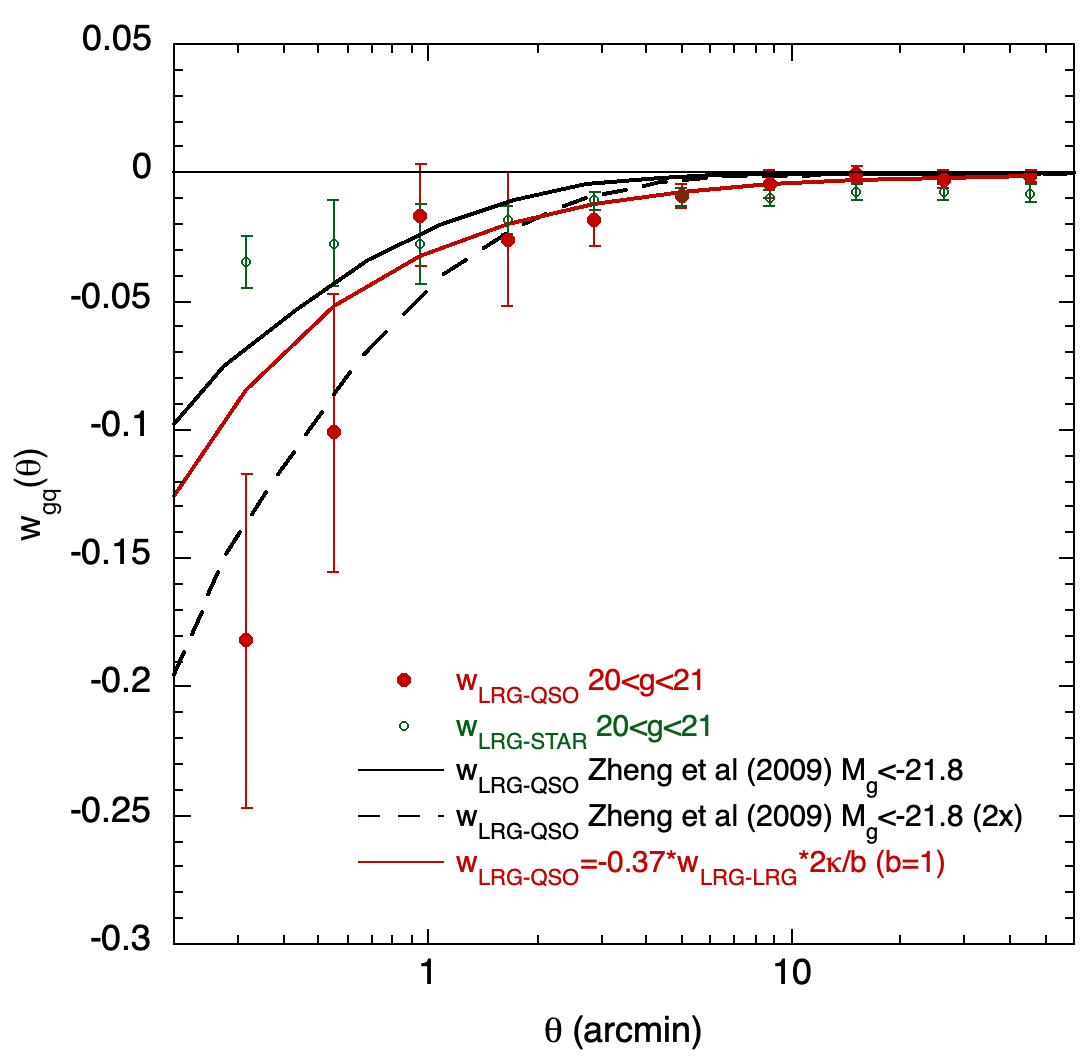}
   \vspace{-0.4cm}
    \caption[LRG HOD models]{(a) $w_{gg}$ and $w_{g\kappa}$ auto- and cross-correlation functions predicted by the HOD model of \cite{Zheng2009} for $M_g<-21.8$ SDSS LRGs (with $\sigma_8=0.8$ and $h=0.7$), compared to $w_{gg}(\theta)$ for our LRGs. (b) The cross-correlation function, $w_{gq}(\theta)$, for $17<g<19$ QSO's and our LRG sample, compared to the  HOD model of \cite{Zheng2009} and the 2 models of \cite{WI1998} with $b=1$ and $b=0.6$. (c) Same as (b) for the QSO magnitude range $20<g<21$. }
    \label{fig:wgm_withsp7HODmodels}
\end{figure}

We next attempt to model the VST ATLAS LRG sample that are assumed to occupy the $0.16<z<0.36$ range with an approximately flat $n(z)$  (see Fig. 12 of \citealt{Eisenstein2001}). From Fig. \ref{fig:wgm_withsp7HODmodels} (a) we see that the LRG auto-correlation function $w_{LRG-LRG}$ is $\approx10\times$ higher than the $17<r<21$ galaxy $w_{gg}$ in Fig. \ref{fig:wgm_withHODmodels} (a). The higher amplitude clustering of the LRGs  will allow more powerful weak lensing tests of the 1- and 2-halo terms for HODs claimed to be appropriate for LRGs. So we shall now test  the LRG HOD model advocated by \cite{Zheng2009}  with $M_g<-21.8$ (see Fig. \ref{fig:HOD_NM_models}) and first compare it to our LRG $w_{gg}(\theta)$ in Fig. \ref{fig:wgm_withsp7HODmodels} (a). While reaching the amplitude of the observed LRG $w_{gg}(\theta)$ at $\theta\approx1'$, we see that the HOD predicted $w_{gg}$ again underestimates the observations at  scales of $\approx10'$, similar to what was found for the $17<r<21$ galaxy HOD model of \cite{ZhengCoilZehavi2007} in Fig. \ref{fig:wgm_withHODmodels} (a).   The fit also appears somewhat worse at large scales than found for the  SDSS LRG $w_p(\sigma)$ by \cite{Zheng2009}.  Nevertheless, since the HOD model fits $w_{gg}$ in the range $\theta<5'$ we again suggest that it is a useful basis to  test the HOD model of \cite{Zheng2009}  against the simpler \cite{WI1998} model using the LRG-QSO cross-correlations as considered in Section \ref{sec:LRG_QSO} below.

\subsection{LRG-QSO lensing}
\label{sec:LRG_QSO}

As before for galaxies, we investigate the mass distribution around LRGs by analysing their cross-correlation with $17<g<19$ and $20<g<21$ ATLAS QSO samples, but based here first on the \cite{Zheng2009} HOD model for $w_{g\kappa}$ as shown by the dashed line in Fig. \ref{fig:wgm_withsp7HODmodels} (a). 
We see that for the $20<g<21$ QSO case in  Fig. \ref{fig:wgm_withsp7HODmodels} (c), a significant anti-correlation signal is seen at $\theta<5'$ and particularly at $\theta\approx0.'3$ where ${w_{gq}}\approx-0.17$, even taking into account that the control star sample also shows a less significant anti-correlation at $\theta\approx0.'3$. However, a less strong signal is seen in the $17<g<19$ case in  Fig. \ref{fig:wgm_withsp7HODmodels} (b) where $w_{gq}$ is consistent with zero at all scales; we note that the errors are larger here. We then checked for the presence of dust by re-doing  the cross-correlations with the QSO samples limited at bright and faint W1 magnitudes. The bright cross-correlation is  expected to increase more than the faint cross-correlation in the case of dust due to the steeper QSO $n(g)$. However, both the bright and faint W1 cross-correlations were consistent with the g-limited results in Figs. \ref{fig:wgm_withsp7HODmodels} (b, c). Inspection of the $w_{gq}$ results in the 8 sub-areas used for the field-field errors also showed that the anti-correlation existed in almost all sub-areas. \\

We compare to the model of \cite{Zheng2009} for $M_g<-21.8$ LRG's (solid black line) and see that although it is consistent with the bright  QSO cross-correlation in  Fig. \ref{fig:wgm_withsp7HODmodels} (b),  it remains  above the less noisy faint QSO result at most scales below $\theta\approx5'$ in  Fig. \ref{fig:wgm_withsp7HODmodels} (c), although the reduced $\chi^2$ is still only 1.73 for these 6 points.  As in Section \ref{sec:lrg_hod_model} above, we have assumed a flat $n(z)$ in the range $0.16<z<0.36$ for the LRGs, following \cite{Eisenstein2001}. A \cite{WI1998} model with $b=1$ based on the LRG-LRG autocorrelation function in  Fig. \ref{fig:wgm_withsp7HODmodels} (a) is also shown in Figs. \ref{fig:wgm_withsp7HODmodels} (b, c) assuming the same optical depth ($\kappa=0.025$) used previously for the $r<21$ galaxy sample in Section~\ref{sec:QSOgal_xcorr}. This model assumes that the LRGs trace the mass and this model does get closer to the $w_{gq}$ results than the above HOD model. However, the low point at $\theta=0.'3$ remains over-estimated by both.  To check if it's the form or the amplitude of the halo mass profile that is causing the problem we show the HOD model multiplied by a factor of 2 as the dashed line in Fig. \ref{fig:wgm_withsp7HODmodels} (c); the fit improves suggesting that it may be the amplitude rather than the form of the NFW mass profile that is at fault. \\

We conclude that the \cite{Zheng2009} LRG HOD that gives a reasonable fit to the ATLAS LRG $w_{gg}$ at least at small, $\theta<2'$ scales may be rejected by $w_{gq}$ in the same angular range. The problem seems to be that the effective bias produced by the HOD appears too small and a higher amplitude mass profile may be needed to improve the fit. We also note that the LRG HOD also underestimates the LRG $w_{gg}$ at larger scales and this might only be addressed by using a higher value of $\sigma_8>\approx1$ which seems another problem for the LRG HOD approach  at larger scales to put alongside the lensing magnification problem at smaller scales.

\subsection{Further LRG-CMB lensing test of HOD model}

\begin{figure}
    \centering
 	\includegraphics[width=\columnwidth]{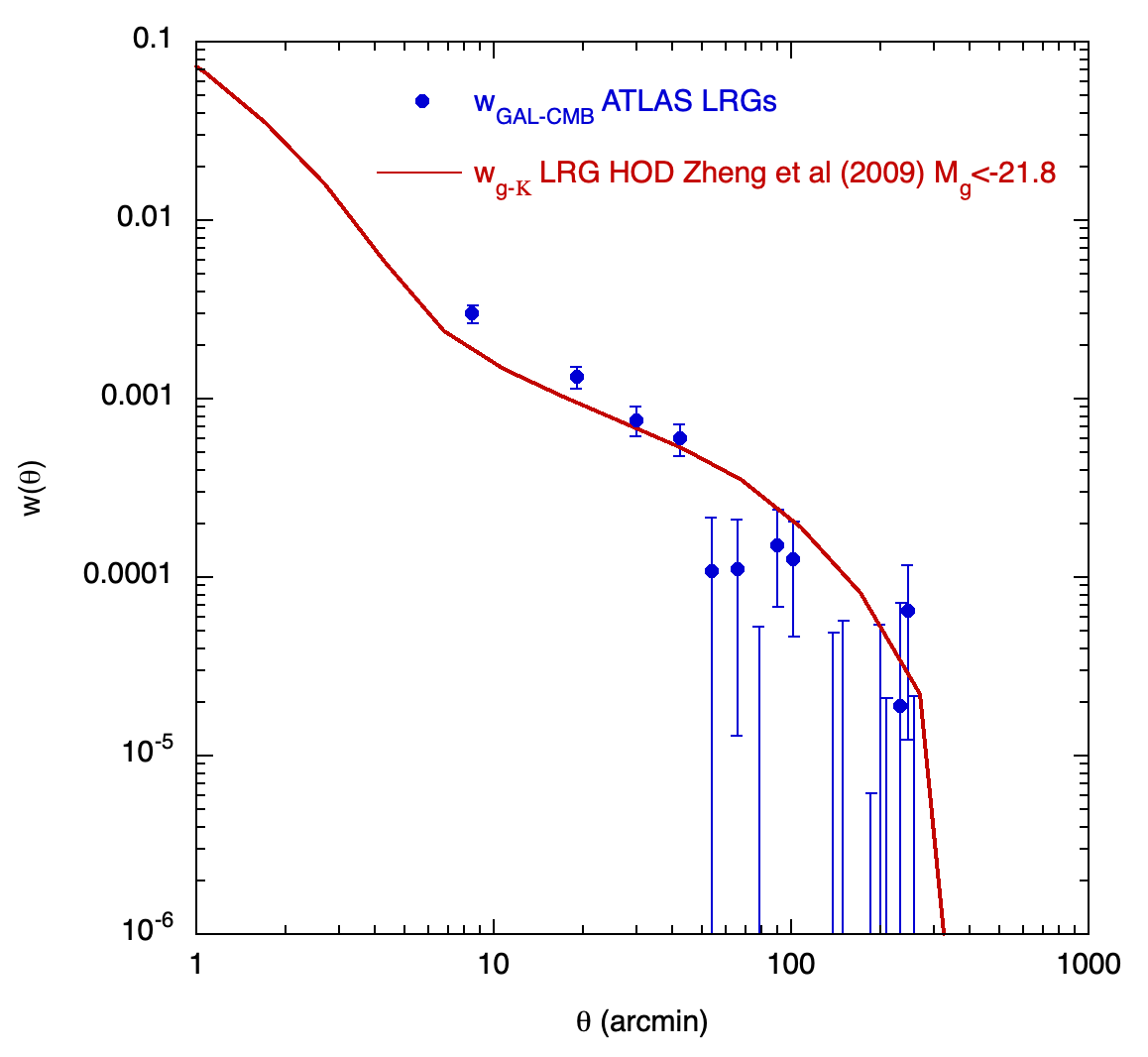} 
	\caption[HOD models.]{ $w_{g-CMB}$  cross-correlation function for our LRG sample with $0.16<z<0.36$ and the Planck (2018) Lensing Map with field-field errors, compared to the HOD model of  \cite{Zheng2009} with parameters $log(M_{min})=14.45$, $\sigma_M=0.80$, $log(M_0)=  12.64$, $log(M_{1'})=15.10$, $\alpha=1.72$.  The model assumes $h=0.7$ and $\sigma_8=0.8$. The first bin centre at 6$'$ corresponds to $\approx1$ h$^{-1}$Mpc at the LRG mean $z$ of $z=0.26$.}
	\label{fig:w_LRGcmb_withHODmodels}
\end{figure}

In Fig. \ref{fig:w_LRGcmb_withHODmodels} we show the $0.16<z<0.36$ LRG cross-correlation with the Planck CMB lensing convergence map compared to the prediction of the \cite{Zheng2009} HOD model. {Overall, the model fits the data well, giving reduced $\chi^2=1.7$. Looking in more detail and in the context of the fit of the Z09 model to the LRG $w_{gg}$ in Fig. \ref{fig:wgm_withsp7HODmodels}(a), we see reasonable agreement between the data and model here for the 2-halo term at $\theta>10'$ although this is the range where the model significantly underpredicts the LRG $w_{gg}$. At smaller scales where the Z09 model fits the LRG $w_{gg}$ very well, the CMB lensing prediction is too low compared to the observed result, in agreement with the LRG-QSO cross-correlation result seen in Fig. \ref{fig:wgm_withsp7HODmodels}(c). Thus the LRG HOD model  either fits the LRG $w_{gg}$ while  underestimating the QSO and CMB lensing results at small scales or underpredicts the LRG $w{gg}$ while  fitting the CMB lensing result at large scales. This is reasonably consistent with the galaxy lensing results in  Section~\ref{sec:Gal_CMB_HOD}, where at small scales the HOD underestimates the  galaxy-QSO cross-correlation $w_{gq}$ relative to the galaxy $w_{gg}$ whereas at large scales the Z07 HOD, at least, fits $w_{gg}$ while over-predicting the galaxy CMB lensing result. However, the LRG results are stronger because of their high amplitude and signal-noise. Similar large scale behaviour may also be seen in the CMB lensing results for the $n>5$ groups and clusters sample in  Fig. \ref{fig:wCK_NFW_Mmp3} and Table \ref{tab:cluster_masses} of Section \ref{sec:cluster_cmb}.

\section{Conclusions}
\label{sec:conclusions}

We have detected lensing magnification of background quasars by foreground clusters, galaxies and LRGs. We have used stars as control samples and these have suggested there may be a sky-subtraction problem for the NEO7 and DECALS DR10 W1 and W2 magnitudes when measured in the vicinity of bright galaxies. We have also investigated lensing of the CMB by  these VST ATLAS cluster and galaxy samples and detected  strong effects in each case.

From the  lensing of ATLAS quasars by galaxy clusters in the ATLAS  catalogue of \cite{Ansarinejad2023} we find that NFW profiles with halo masses  of $\approx1\times10^{15}$M$_\odot$ fit clusters with $n>40$ members with $\approx1\times10^{14}$M$_\odot$ fitting groups/clusters of $n>5$  members. The $n>40$ clusters show the greatest signal but both cluster samples show a preference for an NFW profile over an SIS at the small scales probed by quasar lensing.  The larger scales dominated by the 2-halo terms are much better investigated using CMB lensing. Cross-correlation of the {\it Planck} CMB lensing convergence map with the galaxy clusters showed very strong signals for both cluster samples and we find cluster masses of $\approx1\times10^{14}$M$_\odot$ for the $n>5$ clusters and $\approx3\times10^{14}$M$_\odot$ for the $n>40$ clusters. Overall, the quasar and CMB lensing mass estimates are in good agreement for both samples. However,  the CMB lensing cross-correlation is less well fitted by the  $n>5$ sample than is the $n>40$ sample. Also while the quasar and CMB lensing masses bracket the average  masses quoted for the $n>40$ clusters, the lensing mass estimates for the $n>5$ sample are generally lower than those quoted by \cite{Ansarinejad2023} by a factor of $\approx3-5$.

For the VST ATLAS $17<r<21$ galaxy sample, we find that galaxy-galaxy angular auto-correlation and the quasar - galaxy cross-correlation results are consistent with those for SDSS galaxies by respectively \cite{Wang2013} and \cite{Scranton2005} and both are at similar levels of significance. We then addressed the question of how e.g. \cite{Myers2005} found too high a level of QSO magnification for compatibility with standard $\Lambda$CDM cosmology compared to \cite{Scranton2005} who found that the SDSS QSO magnification studies were consistent with standard cosmology predictions. Generally we agree with the previous conclusions of \cite{MS2007} that the actual observations are very consistent with each other and that the difference lay in the models used to interpret these quasar-galaxy cross-correlations. Previously, \cite{Myers2005} assumed that galaxies traced the mass up to a linear bias factor and we have again shown on this assumption that values of the galaxy bias much smaller than unity or equivalently values of $\sigma_8$ higher than unity are needed for such models to fit. If instead the  HOD approach of \cite{Scranton2005} is followed, then models such as the SDSS $M_r<-20.8$ model of \cite{Zheng2007} can be found that at least approximately fit our measured galaxy angular auto-correlation function while simultaneously reasonably fitting the QSO-galaxy cross-correlation function at the same scales. However, there is a hint that the \cite{Zheng2007} model that fits $w_{gg}$ is still slightly too low in lensing magnification amplitude at the smallest scales of $w_{gq}$. Our strong detection of the ATLAS galaxy- {\it Planck} CMB lensing signal was also slightly over-predicted by the \cite{Zheng2007} HOD model at a similar level as the 2-halo term's over-prediction of the group/cluster $n>5$ sample. But both these deficiencies were only marginally detected and this motivated us to look at the lensing results for the more extreme case of highly clustered LRGs to see if any such problems persisted there. 

We therefore selected a sample of ATLAS $r<19.5$ LRG's, using similar criteria to the SDSS Cut 1 of \cite{Eisenstein2001} with a $0.16<z<0.36$ redshift range and found an LRG auto-correlation function amplitude $\approx10\times$ that of the above $17<r<21$ galaxy sample. We found that the LRG HOD of \cite{Zheng2009} again fitted $w_{gg}(\theta)$ well at small scales but underestimated $w_{gg}$ at larger scales, similar to the galaxy HOD. We then compared the LRG HOD prediction to the QSO-LRG cross-correlation function and found that it under-predicted the amplitude of the LRG anti-correlation with $20<g<21$ ATLAS quasars, at a level stronger than the hint in the $17<r<21$ galaxy $w_{gq}$. Multiplying the LRG HOD prediction by a factor of 2 significantly improved the fit, demonstrating the size of the effect. The $17<g<19$ QSO-LRG cross-correlation showed less discrepancy with the HOD prediction but here the errors are much larger.

Overall, we conclude that our QSO-galaxy cross-correlation results are in good agreement with previous authors for clusters and $17<r<21$ galaxies and that HOD models improve standard $\Lambda$CDM cosmology fits, in particular in the $17<r<21$ galaxy case compared to models where galaxies trace the mass. In the case of  clusters,  NFW mass profiles are preferred over SIS profiles, with NFW mass estimates compatible with previous results for both clusters and groups. CMB lensing results for groups tended to be under-predicted by standard 2-halo models and this was also seen marginally in the CMB lensing of the $17<r<21$ galaxies.   LRGs show the biggest discrepancies with a standard HOD model, where they under-predict $w_{gq}$ by a factor of $\approx2\times$ in the fainter QSO samples, while over-predicting the LRG-CMB lensing result by a smaller factor. Further investigation is required to see if improved HOD models can be found to address these anomalies at large and small scales in the galaxy, group and particularly LRG samples.
    
\section*{Data Availability Statement}
The ESO VST ATLAS and WISE    data we have used are all publicly available. The VST ATLAS QSO Catalogue can be found at \url{https://astro.dur.ac.uk/cea/vstatlas/qso_catalogue/}. All other data relevant to this publication will be supplied on request to the authors.

\section*{Acknowledgements}

We acknowledge use of the ESO VLT Survey Telescope (VST) ATLAS. The ATLAS survey is based on data products from observations made with ESO Telescopes at the La Silla Paranal Observatory under program ID 177.A-3011(A,B,C,D,E.F,G,H,I,J,K,L,M,N) (see \citealt{Shanks2015}).

We acknowledge the use of  data products from WISE, which is a joint project of the University of California, Los Angeles, and the Jet Propulsion Laboratory (JPL)/California Institute of Technology (Caltech), funded by the National Aeronautics and Space Administration (NASA), and from NEOWISE, which is a JPL/Caltech project funded by NASA. 

We acknowledge use of SDSS imaging and spectroscopic data. Funding for SDSS-III has been provided by the Alfred P. Sloan Foundation, the Participating Institutions, the National Science Foundation and the US Department of Energy Office of Science. 

BA acknowledges support from the Australian Research Council’s Discovery Projects scheme (DP200101068).

LFB acknowledges support from ANID BASAL project FB210003.

We finally acknowledge STFC Consolidated Grant  ST/T000244/1 in supporting this research.

For the purpose of open access, the author has applied a Creative Commons Attribution (CC BY) licence to any Author Accepted Manuscript version arising.

We thank the referee for useful comments that have improved the paper.



\bibliographystyle{mnras}
\bibliography{bibliography} 









\bsp	
\label{lastpage}
\end{document}